\newif\ifAMStwofonts
\AMStwofontstrue

\def\lsim{ \lower .75ex \hbox{$\sim$} \llap{\raise .27ex \hbox{$<$}} }
\def\gsim{~\rlap{$>$}{\lower 1.0ex\hbox{$\sim$}}}

\def\dd{{\rm d}}

\def\HII{H{\sc ii}}
\def\HI{H{\sc i}}

\documentstyle[psfig,mnras_cite]{mn}
\def\fesctxt{$f_{\rm esc}$}

\title{Non-Uniform Reionization by Galaxies and
its Effect on the Cosmic Microwave Background}
\author[A.~J.~Benson, Adi Nusser, Naoshi Sugiyama and C.~G.~Lacey]{A.~J.~Benson$^{1,5}$, Adi Nusser$^2$, Naoshi Sugiyama$^3$ and C.~G.~Lacey$^{1,4}$  \\
$^1$ Department of Physics, University of Durham, UK.\\
$^2$ The Physics Department, The Technion-Israel Institute of Technology,
Technion City, Haifa 32000, Israel.\\
$^3$ Department of Physics, Kyoto University, Kyoto 606-8502, Japan.\\
$^4$ SISSA, via Beirut, 2-4, 34014 Trieste, Italy.\\
$^5$ E-mail: A.J.Benson@dur.ac.uk\\}

\begin{document}

\maketitle

\begin{abstract}
We present predictions for the reionization of the intergalactic
medium (IGM) by stars in high-redshift galaxies, based on a
semi-analytic model of galaxy formation. We calculate ionizing
luminosities of galaxies, including the effects of absorption by
interstellar gas and dust on the escape fraction $f_{\rm esc}$, and
follow the propagation of the ionization fronts around each galaxy in
order to calculate the filling factor of ionized hydrogen in the
IGM. For a $\Lambda{\rm CDM}$ cosmology, with parameters of the galaxy
formation model chosen to match observations of present-day galaxies,
and a physical calculation of the escape fraction, we find that the
hydrogen in the IGM will be reionized at redshift $z=6.1$ if the IGM
has uniform density, but only by $z=4.5$ if the IGM is clumped.  If
instead we assume a constant escape fraction of 20\% for all galaxies,
then we find reionization at $z=9.0$ and $z=7.8$ for the same two
assumptions about IGM clumping. We combine our semi-analytic model
with an N-body simulation of the distribution of dark matter in the
universe in order to calculate the evolution of the spatial and
velocity distribution of the ionized gas in the IGM, and use this to
calculate the secondary temperature anisotropies induced in the cosmic
microwave background (CMB) by scattering off free electrons. The
models predict a spectrum of secondary anisotropies covering a broad
range of angular scales, with fractional temperature fluctuations
$\sim 10^{-7}-10^{-6}$ on arcminute scales. The amplitude depends
strongly on the total baryon density, and less sensitively on the
escape fraction $f_{\rm esc}$. The amplitude also depends somewhat on
the geometry of reionization, with models in which the regions of
highest gas density are reionized first giving larger CMB fluctuations
than the case where galaxies ionize surrounding spherical regions, and
models where low density regions reionize first giving the smallest
fluctuations. Measurement of these anisotropies can therefore put
important constraints on the reionization process, in particular, the
redshift evolution of the filling factor, and should be a primary
objective of a next generation submillimeter telescope such as the
Atacama Large Millimeter Array.
\end{abstract}


\section{Introduction}

The Gunn-Peterson (GP) effect \cite{gunn65} strongly indicates
that the smoothly distributed hydrogen in the intergalactic medium
(IGM) is already highly ionized by $z=5$
\cite{schneider91,lanzetta95}. Barring the possibility of
collisional reionization (e.g. Giroux \& Shapiro 1994), the GP effect
implies the presence of very luminous ionizing sources at high
redshifts capable of producing enough Lyman continuum (Lyc) photons to
cause photoionization of hydrogen by $z\gsim 5$. The two possible
sources of these ionizing photons are QSOs and high mass stars.

Models in which QSOs dominate the production of ionizing photons may
be able to meet the GP constraint \cite{miralda90}. However, such
models are strongly constrained by the observed drop in the abundance
of bright QSOs above $z\approx 3$
\cite{hartwick90,warren94,kennefick95,schmidt95}. Furthermore,
\scite{mhr98} note that a model in which faint QSOs provide all
the required ionizing luminosity can be ruled out on the basis of the
number of faint QSOs seen in the HDF.

There is, however, growing evidence for the presence of bright
galaxies at redshifts as high as $z \sim 5$ \cite{spinrad98}, and
perhaps even higher \cite{yahil98}. Thus the other natural candidate
sources of ionizing photons are young, high mass stars forming in
galaxies at redshifts greater than 5
(e.g. \pcite{CR86,haiman96,ciardi99}). \scite{mhr98} note that at
$z\approx 3$ stars in Lyman-break galaxies will emit more ionizing
photons into the IGM than QSOs if more than 30\% of such photons can
escape from their host galaxy. Whilst such high escape fractions may
not be realistic (e.g. local starbursts show escape fractions of only
a few percent, \pcite{leitherer95}), this does demonstrate that
high-redshift galaxies could provide a significant contribution to (or
perhaps even dominate) the production of ionizing photons. In this
work we will restrict our attention to ionizing photons produced by
stars, deferring consideration of the QSO contribution to a later
paper.

According to the hierarchical structure formation scenario
(e.g. \pcite{peebles80}) perturbations in the gravitationally dominant
and dissipationless dark matter grow, by gravitational instability,
into virialised clumps, or halos.  Galaxies, and later stars, then
form by the cooling and condensation of gas inside these halos
(e.g. \pcite{wr78,wf91}). Dark matter halos continually grow by
merging with other halos (e.g. \pcite{bower91,bcek}). In the context
of this hierarchical scenario, we present a realistic scheme for
studying the reionization of the universe by ionizing photons emitted
from massive stars.  We focus on the photoionization of the hydrogen
component of the IGM. To predict the time dependent luminosity in Lyc
photons we use a semi-analytic model of galaxy formation
(e.g. \pcite{kauff93,coleetal94,somerville98}). In
particular, we use the semi-analytic model of \scite{coleetal99},
modified to take into account Compton cooling by cosmic microwave
background (CMB) photons, to model the properties of galaxies living in
dark matter halos spanning a wide range of masses.

We then estimate the fraction of the ionizing photons which manage to
escape each galaxy, and therefore contribute to the photoionization of
the intergalactic \HI. The fraction of ionizing photons escaping is
determined on a galaxy-by-galaxy basis, using physically motivated
models. Assuming spherical symmetry, we follow the propagation of the
ionization front around each halo to compute the filling factor of
intergalactic \HII\ regions, including the effects of clumping in the
IGM.  Finally, using several alternative models for the spatial
distribution of ionized regions within a high resolution N-body
simulation of the dark matter distribution, we estimate the
anisotropies imprinted on the CMB by the patchy reionization process,
due to the correlations in the ionized gas distribution and velocities
\cite{sz80,vishniac87}. In previous models many simplifications were
made in computing both the spatial distribution of ionized regions and
the two-point correlations of gas density and velocity in those
regions
\cite{aghanim96,jaffe98,gruzinov98,knox98,peebles98,haiman99}. Our
calculations represent a significant improvement over these models as
we are able to calculate the two-point correlations between gas
density and velocity in ionized regions directly from an N-body
simulation.

The rest of this paper is arranged as follows. In
\S\ref{sec:modified} we outline the features of the semi-analytic
model relevant to galaxy formation at high redshifts. In
\S\ref{sec:fescape} we describe how we calculate the fraction of
ionizing photons escaping from galaxies, and observational constraints
on the ionizing luminosities and escape fraction at low and high
redshift from ${\mathrm H}\alpha$ luminosities and \HI\ masses and
column densities.  In \S\ref{sec:fronts} we describe how we calculate
the filling factor of photoionized gas in the IGM, including the
effects of clumping of this gas. We then present our predictions for
reionization, including the effects on the reionization redshift of
using different assumptions about escape fractions and clumping
factors.In \S\ref{sec:variants} we examine the robustness of our
results to changes in the other parameters of the semi-analytic galaxy
formation model. In \S\ref{sec:spatial} we describe how the
semi-analytic models are combined with N-body simulations to calculate
the spatial distribution of the photoionized IGM. We then calculate
the spectrum of anisotropies introduced into the CMB by this ionized
gas.  Finally, in \S\ref{sec:conc} we summarize our results and
examine their consequences.

\section{The semi-analytic model of galaxy formation}
\label{sec:modified}

To determine the luminosity in ionizing Lyc photons produced by the
galaxy population, we use the semi-analytic model of galaxy formation
developed by \scite{coleetal99}. This model predicts the
properties of galaxies residing within dark matter halos of different
masses.  This is achieved by relating, in a self-consistent way, the
physical processes of gas cooling, star formation, and supernovae
feedback to a halo's merger history, which is calculated using the
extended Press-Schechter theory. The parameters of this model are
constrained by a set of observations of galaxies in the local
Universe, including the B and K-band luminosity functions, the I-band
Tully-Fisher relation, the mixture of morphological types and the
distribution of disk scale lengths (see
\scite{coleetal99} and references therein for a thorough
discussion of the observational constraints). Once the model has been
constrained in this way it is able to make predictions concerning the
clustering of galaxies \cite{ajbclust} and the properties of galaxies
at higher redshifts. For reference, the parameters of our standard
model are given in Table~\ref{tb:standpars}. Definitions of the
semi-analytic model parameters can be found in
\scite{coleetal99}.

\begin{table}
\caption{The parameters of our standard model.}
\label{tb:standpars}
\begin{center}
\begin{tabular}{ll}
\hline
\textbf{Parameter} & \textbf{Value} \\
\hline
\multicolumn{2}{c}{{\em Cosmology}}\\
\hline
$\Omega_0$ & 0.3 \\
$\Lambda_0$ & 0.7 \\
$H_0$ & 70 km/s/Mpc \\
$\sigma_8$ & 0.90 \\
$\Gamma$ & 0.21 \\
$\Omega_{\rm b}$ & 0.02 \\
\hline
\multicolumn{2}{c}{{\em Gas cooling}}\\
\hline
Gas profile$^\P$ & CDC \\
$r_{\rm core}$ & $0.33 r_{\rm NFW}$ \\
Recooling$^*$ & not allowed \\
\hline
\multicolumn{2}{c}{{\em Star formation and feedback}}\\
\hline
$\alpha_\star$ & -1.5 \\
$\epsilon_\star$ & 0.01 \\
$\alpha_{\rm hot}$ & 2.0 \\
$V_{\rm hot}$ & 150 km s$^{-1}$ \\
Feedback$^{**}$ & standard \\
\hline
\multicolumn{2}{c}{{\em Stellar populations}}\\
\hline
IMF & \scite{kennicutt83} \\
$p$ & 0.02 \\
$R$ & 0.31 \\
$\Upsilon$ & 1.53 \\
\hline
\multicolumn{2}{c}{{\em Mergers and bursts}}\\
\hline
$f_{\rm df}$ & 1.0 \\
$f_{\rm ellip}$ & 0.3 \\
Starbursts & included \\
$f_{\rm dyn}$ & 1.0 \\
\hline
\multicolumn{2}{c}{{\em Ionizing luminosities}}\\
\hline
$S_2^{\ddag}$ & $10.0 \times 10^{50}$ photons/s \\
$h_{\rm z}/r_{\rm disk}^\dag$ & 0.1 \\
\hline
\end{tabular}
\end{center}
\P\ Gas profiles that we consider are CDC (for which the gas density is $\rho (r) \propto (r^2+r_{\rm core}^2)^{-1}$ --- i.e. an isothermal profile but
with a constant density core), and SIS (an isothermal profile with no
core).\\

$*$ Gas ejected from galaxies by supernovae is allowed to cool again
in the same halo if `Recooling' is allowed. Otherwise this gas can
only cool again once it enters a newly formed halo.\\

$**$ `Standard' feedback is the form specified in
eqn. (\ref{eq:beta}). The alternative is `modified' feedback, which is
the form specified in eqn. (\ref{eq:modfb}).\\

$\ddag$ The upper cut off the in luminosity function of OB
associations assumed in the DS94 model (see Appendix~\ref{append:fesc}
and \S\ref{sec:fescmodels}).\\

$\dag$ This is the ratio of disk vertical and radial scale lengths
used in the DS94 and DSGN98 models for the gas escape fraction (see
Appendix \ref{append:fesc} and \S\ref{sec:fescmodels}).\\

\end{table}

As we are employing the semi-analytic model at much higher redshifts
than we have previously attempted, we will investigate the effects on
our results of changing key model parameters. Of particular interest
will be the prescription for feedback from supernovae and stellar
winds. The model assumes that a mass $\beta\Delta M$ of gas is
reheated by supernovae and ejected from the disk for each mass $\Delta
M$ of stars formed. The quantity $\beta$ is allowed to be a function
of the galaxy properties, and is parameterised as
\begin{equation}
\beta = (V_{\mathrm disk}/V_{\mathrm hot})^{-\alpha _{\mathrm hot}},
\label{eq:beta}
\end{equation}
where $V_{\mathrm disk}$ is the circular velocity of the galaxy disk
and $V_{\mathrm hot}$ and $\alpha _{\mathrm hot}$ are adjustable
parameters of the model. \scite{coleetal99} show that $\alpha
_{\mathrm hot}$ and $V_{\rm hot}$ are well constrained by the shape of
the B-band luminosity function and the Tully-Fisher relation at
$z=0$. However, since there is very little time available for star
formation at the high redshifts which we consider, it is possible that
these parameters, or indeed the form of the parameterisation in
eqn. (\ref{eq:beta}), could be changed at high redshift without
significantly affecting the model predictions at $z=0$. In
\S\ref{sec:variants} we will therefore experiment with different
values of these parameters and will also consider a modified
functional form for $\beta$.

Two other key model inputs are the baryon density parameter, $\Omega
_{\rm b}$, and the stellar initial mass function (IMF). The value of
$\Omega _{\mathrm b}$ determines cooling rates (and so star formation
rates) in our model halos. The shape of the IMF determines the number
of high mass stars which produce the ionizing photons. For $\Omega
_{\rm b}$ our standard value is 0.02, which is consistent with the
estimate \scite{walker91} and which allows a good match to the
bright end of the observed B-band luminosity function. We will also
consider an alternative value of $\Omega_{\rm b}=0.04$, which is in
better agreement with estimates from the D/H ratio in QSO absorption
line systems \cite{schramm98,burles98}. For the IMF, we adopt as our
standard choice the IMF of \scite{kennicutt83}, which is close to
the ``best'' IMF proposed by \scite{scalo98} on the basis of
observations in the Solar neighbourhood and in nearby galaxies. We
consider the effects of changing both $\Omega_{\rm b}$ and the IMF in
\S\ref{sec:variants}.

\subsection{Gas Cooling}
\label{sec:cooling}

The standard semi-analytic model of \scite{coleetal99} allows hot
halo gas to cool only via collisional radiative processes. At high
redshift, Compton cooling due to free electrons in the hot plasma
scattering off CMB photons becomes important. The Compton cooling
timescale is given by \cite{peebles68}
\begin{equation}
t_{\mathrm Compton} = {1161.3 (1 + x_{\mathrm hot}^{-1}) \over (1+z)^4
(1 - T_0^{\mathrm CMB} (1+z)/T_{\mathrm e})} \hbox{Gyr},
\label{eq:compton}
\end{equation}
where $x_{\mathrm hot}$ is the ionized fraction of the hot halo gas,
$T_0^{\mathrm CMB}$ is the temperature of the CMB at the present day
and $T_{\mathrm e}$ is the temperature of electrons in the hot halo
gas, which we set equal to the virial temperature of the halo. At high
redshifts this cooling time becomes shorter than the Hubble time and
so Compton cooling may be effective at these redshifts. To implement
eqn. (\ref{eq:compton}) in the semi-analytic model, we assume that the
shock-heated halo gas is in collisional ionization equilibrium, and
use values for $x_{\mathrm hot} $ which we interpolate from the
tabulated values given by \scite{sutherland93}. In halos with
virial temperatures less than around $10^4$K collisional ionization is
ineffective, and so the ionized fraction in the halo gas will equal
the residual ionization fraction left over from
recombination. However, since this fraction is small, we will simply
assume in this paper that cooling in halos below $10^4$K is
negligible.

It should be noted that, unlike the radiative cooling time, the
Compton cooling time is independent of the gas density and depends
only very weakly on the gas temperature. Whereas with collisional
radiative cooling a cooling radius, within which the cooling time is
less than the age of the halo, propagates through the halo as more and
more gas cools, with Compton cooling the entire halo cools at the same
rate. The amount of gas which can reach the centre of the halo is then
controlled by the free-fall timescale in the halo.

Including Compton cooling in our model turns out to make little
difference to the results. For example, the total mass of stars formed
in the universe as a function of redshift differs by less than $5\%$
for $z<20$ between models with and without Compton cooling. At higher
redshifts the differences can become as large as 30\% for small
intervals of redshift. For example, if a massive halo cools via
Compton cooling it will rapidly produce many stars --- without Compton
cooling it will still form these stars, but not until slightly later
when collisional radiative cooling takes effect. However, the mass of
stars formed at these high redshifts is tiny, and so any differences
become entirely negligible at lower redshifts when many more stars
have formed. Although at high redshift the Compton cooling time is
shorter than the age of the Universe, halos are merging at a high rate
and so their gas is being repeatedly shock-heated by successive
mergers which, we assume, heat the gas to the virial temperature of
the halo. We find that for the majority of halos the time between
successive major mergers (defined as the time for a doubling in mass
of the halo) is less than the Compton cooling time at the redshifts
considered here. Therefore, Compton cooling will be ineffective in
these halos. In the few cases where the halo does survive long enough
that Compton cooling could be important, we find that the collisional
radiative cooling time at the virial radius of the halo is often
shorter than the Compton cooling time, in which case all of the gas in
the halo will cool whether or not we include the effects of Compton
cooling. Nevertheless, Compton cooling is included in all models
considered.

In this work we ignore cooling due to molecular hydrogen (H$_2$).
Although molecular hydrogen allows cooling to occur in gas below
$10^4$K, it is easily dissociated by photons from stars that form from
the cooling gas. Previous studies that have included cooling due to
H$_2$ typically find that it is completely dissociated at very high
redshifts. For example, \scite{ciardi99} find that molecular
hydrogen is fully dissociated by $z\approx 25$. Objects formed by
H$_2$ cooling are therefore not expected to contribute significantly
to the reionization of the IGM.

\subsection{Fraction of Gas in the IGM}
\label{sec:IGM}

\begin{figure}
\psfig{file=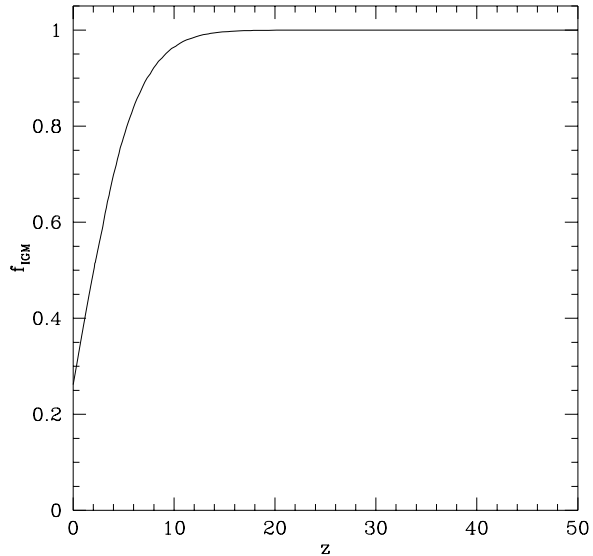,width=80mm} \caption{The fraction of baryons
remaining in the IGM as a function of redshift. Other baryons have
either been collisionally ionized in dark matter halos or have
cooled to become part of a galaxy.} \label{fig:fIGM}
\end{figure}

At any given redshift, some fraction of the gas in the Universe will
have become collisionally ionized in dark matter halos and some
fraction will have cooled to become part of a galaxy. Within the
context of our semi-analytic model, we define the IGM as all gas which
has \emph{not} been collisionally ionized inside dark matter halos and
which has \emph{not} become part of a galaxy (note that we are here
only interested in ionization of hydrogen). It is this gas which must
be photoionized if the Gunn-Peterson constraint is to be
satisfied. The fraction of the total baryon content of the universe
which is in the IGM, $f_{\rm IGM}$, can be estimated by integrating
over the mass function of dark matter halos, as follows
\begin{equation}
f_{\rm IGM}(z)=1 - \left[ \int_0^\infty M_{\rm gas} x_{\rm H} {{\rm
d}n\over{\rm d}M_{\rm halo}} {{\rm d}M_{\rm halo} \over \Omega_{\rm
b}\rho_{\rm c}} \right] - f_{\rm galaxy}(z),
\label{eq:fIGM}
\end{equation}
where $M_{\rm gas}(M_{\rm halo},z)$ is the mean mass of diffuse gas in
halos of mass $M_{\rm halo}$, $x_{\rm H}(M_{\rm halo},z)$ is the
fraction of hydrogen which is collisionally ionized at the halo virial
temperature (which we take from the calculations of
\pcite{sutherland93}), ${\rm d}n/{\rm d}M_{\rm halo}(M_{\rm halo},z)$
is the comoving number density of halos (which we approximate by the
Press-Schechter mass function), $\rho_{\rm c}$ is the critical density
of the Universe at $z=0$, and $f_{\rm galaxy}$ is the fraction of the
total baryonic mass in the Universe which has been incorporated into
galaxies. The quantities $M_{\rm gas}$ and $f_{\rm galaxy}$ can be
readily calculated from our model of galaxy
formation. Fig. \ref{fig:fIGM} shows the evolution of $f_{\rm IGM}$
with redshift.

\subsection{Observational Constraints}

The semi-analytic model provides the spectral energy distribution
(SED) of each galaxy, from which we can determine the ionizing
luminosity of that galaxy. Summing the contributions from all galaxies
in a given halo yields the total ionizing luminosity produced in that
halo. \scite{cojazzi99}, using the model of
\scite{haiman96}, demonstrated that a higher reionization
redshift could be obtained if zero-metallicity stars were responsible
for reionization, as these produce a greater number of ionizing
photons than low (i.e. $10^{-4}$) metallicity stars.  In our model the
very first stars have zero metallicity, but as we include chemical
evolution only a very small fraction of stars have metallicities below
$10^{-4}$. This is consistent with the results of
\scite{tumlinson99} who argue that the epoch of metal-free star
formation must end before $z=3$, as the enhanced emission shortwards
of 228\AA\ from such stars is inconsistent with observations of He{\sc
ii} opacity in the IGM at that redshift. Therefore we cannot appeal to
such zero-metallicity stars to increase the redshift of reionization
in our model.

As a result of absorption by neutral hydrogen close to the emitting
stars and extinction caused by dust, only a small fraction of
the ionizing radiation emitted by the stars escapes from each galaxy
\cite{leitherer95,hurwitz97,kunth98}. We therefore estimate, within
the context of the semi-analytic model, the fraction, \fesctxt, of
ionizing photons which escape the galaxy to become available for the
photoionization of the \HI\ in the IGM. The calculation of \fesctxt\
is discussed in \S\ref{sec:fescape}.

\begin{figure}
\psfig{file=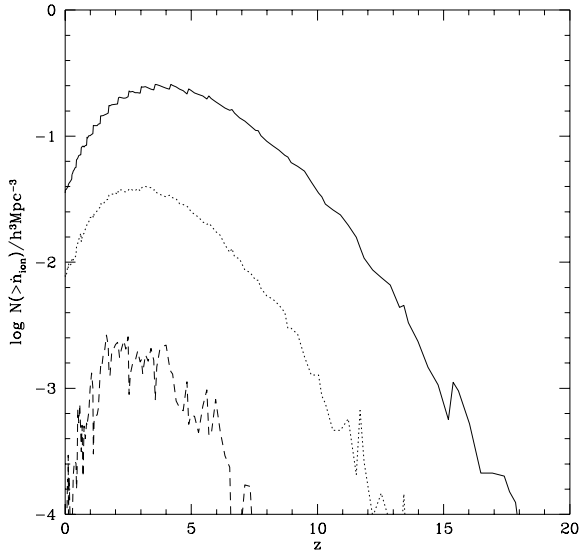,width=80mm}
\caption{The comoving number density of galaxies brighter than a given
ionizing luminosity $\dot{n}_{\mathrm ion}$ as a function of
redshift. The number is plotted for $\dot{n}_{\mathrm ion} = 10^2$
(solid line), $10^3$ (dotted line) and $10^4$ (dashed line) in units of
$10^{50}$ photons/s.}
\label{fig:Ngal}
\end{figure}

In Fig. \ref{fig:Ngal} we show the redshift evolution of the comoving
number density of galaxies with ionizing luminosity $\dot{n}_{\mathrm
ion}$ larger than $10^2$, $10^3$ and $10^4$ in units of $10^{50}$
photons per second. These are the unattenuated luminosities produced
by massive stars in the galaxies. The abundances of sources of given
luminosity rises sharply up to $z=2-4$ (the exact position of the peak
depending on luminosity) as more and more dark matter halos form that
are capable of hosting bright galaxies. After $z=2-4$ abundances
quickly drop towards $z=0$ as the amount of gas available for star
formation declines.

\begin{figure}
\psfig{file=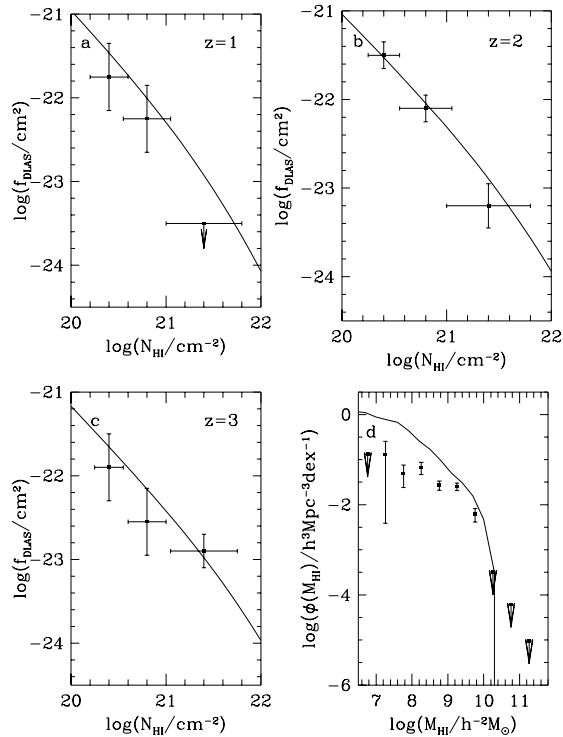,width=80mm}
\caption{Panels (a), (b) and (c) show the distribution of DLAS at
$z=1$, 2 and 3 respectively as a function of their \HI\ column
density. Solid lines indicate the distribution determined from our
model. Points with error bars are from
\protect\scite{lanzetta95}. Panel (d) shows the \HI\ mass
function of galaxies at $z=0.0$. The solid line is the mass function
determined from our model. Points with errorbars are from
\protect\scite{zwaan97}.} 
\label{fig:DLA}
\end{figure}

The escape fractions in our model will be determined by the mass and
radial scale length of the \HI\ gas in galactic disks. It is therefore
important to test that our model produces galaxies with reasonable
distributions of \HI\ mass and disk scale
length. \scite{coleetal99} have shown that our semi-analytic
model produces distributions of I-band disk scale lengths in good
agreement with the $z=0$ data of \scite{dejong99}. In
Fig. \ref{fig:DLA} we compare our model with observations of damped
Lyman-$\alpha$ systems (DLAS) over a range of redshifts and with the
\HI\ mass function at $z=0.0$.  Under the assumption that DLAS are
caused by neutral gas in galactic disks, we compute the DLAS column
density distribution in our model, $f_{\rm DLAS}$, defined such that
$f_{\rm DLAS}(N_{\rm H{\sc i}},t)\dd N_{\rm H{\sc i}} \dd X$ is the
mean number of DLAS at cosmic time $t$ with column densities in the
range $N_{\rm H{\sc i}}$ to $N_{\rm H{\sc i}}+\dd N_{\rm H{\sc i}}$
and absorption distance $X(z) = \frac{2}{3} \left[
(1+z)^{3/2}-1\right]$ in the interval $\dd X$ along a line of sight
\cite{lanzetta95}. Our model is in reasonable agreement with the
distribution of DLAS column densities observed by
\scite{lanzetta95}, indicating that both the mass of \HI\ and its
radial scale length in our model galaxies are realistic. The $z=0$
\HI\ mass function from our model is also in reasonable agreement with
the data of \scite{zwaan97}, although it does overpredict the
abundance of low \HI\ mass galaxies.  Our model predictions assume
that all of the hydrogen in galactic disks is in the form of \HI. In
practice, some of the hydrogen in disks will be in the form of
molecules (H$_2$) or ionized gas (\HII), so this over-estimates the
\HI\ masses and column densities.

A significant contribution to the ionizing luminosity comes from very
low mass halos. We therefore ensure that we resolve all halos which
have a virial temperature $\ge 10^4$K up to $z=50$, i.e. all halos
down to a mass of $5 \times 10^6 h^{-1} M_\odot$. Below this
temperature cooling becomes inefficient (since we are ignoring cooling
by molecular hydrogen, and the Compton cooling from the residual free
electrons left over after recombination) and so galaxy formation
ceases.

The requirement that $5 \times 10^6 h^{-1} M_\odot$ halos be resolved
sets an upper limit on the mass of halo that we can simulate due to
computer memory limits, since the lower the mass of halo that is
resolved, the more progenitors a halo of given mass will have. At
$z=0$, the most massive halos that we are able to simulate make a
significant contribution to the total filling factor and ionizing
luminosity. However, for $z\approx 2$ the most massive halos simulated
contribute only 1\% of the total number of escaping ionizing photons,
and this fraction drops extremely rapidly as we look to even higher
redshifts. Therefore, at the high redshifts ($z\gsim 3$) we will be
interested in, ignoring higher mass halos makes no significant
difference to our results.

We note that once any halo has begun to ionize the surrounding IGM, it
could potentially influence the process of galaxy formation in nearby
halos. Ionizing photons from the first halo will act to heat the gas
in nearby halos, thereby reducing the effective cooling rate
\cite{efstathiou92,thoul96}. Since prior to full reionization each
halo will see only the flux of ionizing photons from nearby sources, a
detailed accounting of this radiative feedback requires a treatment of
the radiative transfer of the ionizing radiation through the IGM. This
is beyond the scope of the present work. Such radiative feedback is
expected to be very efficient at dissociating molecular hydrogen, with
\scite{ciardi99} finding that H$_2$ is completely dissociated by
$z\approx 25$. Radiative feedback will also inhibit galaxy formation
both by reducing the amount of gas that accretes into low mass halos
\cite{gnedin00} and by reducing the cooling rate of gas within
halos. \scite{thoul96} show that radiative feedback may be
effective in inhibiting galaxy formation in halos with circular
velocities of 50 km/s or less. In our model, the ionizing luminosity
becomes dominated by galaxies in halos with circular velocities
greater than 50 km/s at redshifts below $z\approx 10$. At higher
redshifts we may therefore be overestimating the total ionizing
luminosity produced by galaxies, but this should not significantly
affect the reionization redshift.

\section{The escape fraction of ionizing photons}
\label{sec:fescape}

\subsection{Global constraints at low redshift}

Gas and dust inside galaxies can readily absorb ionizing photons and
re-emit the energy at longer wavelengths. Therefore the amount and
distribution of these components are the main factors that determine
\fesctxt . The model of galaxy formation explicitly provides the mass
and metallicity of cold gas present in each galaxy disk and the
half-mass radius of that disk, all as functions of time. The mass of
dust is assumed to be proportional to the mass of cold gas and to its
metallicity.  We split the escape fraction into contributions from
gas, $f_{\rm esc,gas}$, and dust, $f_{\rm esc,dust}$, such that the
total escaping fraction is given by $f_{\rm esc}=f_{\rm esc,gas}f_{\rm
esc,dust}$.

\scite{coleetal99} describe in detail how the effects of dust are
included in their model of galaxy formation. This modelling, which
uses the calculations of \scite{ferrara99}, is much more
realistic than has been previously included in semi-analytic galaxy
formation models, as it includes a fully 3D (though axisymmetric) dust
distribution, and the dust optical depths are calculated for each
galaxy individually. In this model, stars are assumed to be
distributed in a bulge and in an exponential disk with a vertical
scale height equal to $0.0875$ times the radial scale length (this
ratio was adopted by Ferrara et al. to match the observed values for
the old disk population of galaxies like the Milky Way). The dust is
assumed to be distributed in the same way as the disk stars. The
models give the attenuation of the ionizing radiation as a function of
the inclination angle at which a galaxy is viewed, and we average this
over angle to find the mean dust extinction for each galaxy. The dust
attenuations do not include the effects of clumping of the stars or
dust, and also assume that the ionizing stars have the same vertical
distribution as the dust. With these two caveats in mind, the dust
extinctions we apply should only be considered as approximate.

Some of the emitted Lyc photons are absorbed by neutral hydrogen close
to the emitting star, thereby causing H$\alpha$ line emission from the
galaxy. Therefore, the H$\alpha$ luminosity function is sensitive to
the fraction, $f_{\rm esc,gas}$, of the ionizing photons which manage
to escape through the gas. We will require our models to reproduce the
observed H$\alpha$ luminosity function and luminosity density.

\begin{figure}
\psfig{file=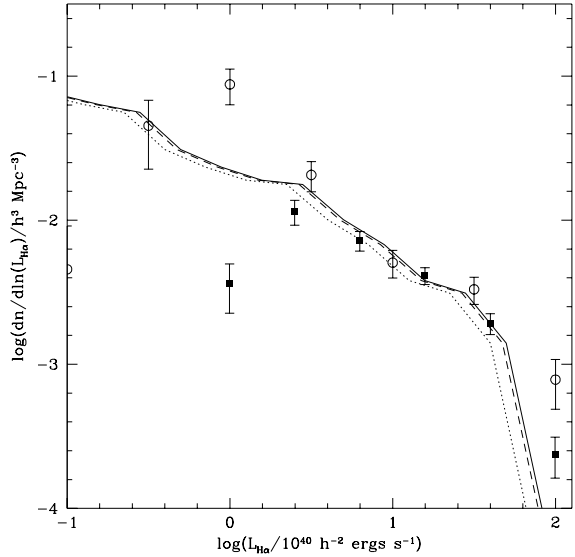,width=80mm}
\caption{The H$\alpha$ luminosity function at $z=0$. Points with error
bars are observational data from \protect\scite{gallego95}
(filled squares) and \protect\scite{sullivan99} (open circles;
note that the median redshift of this survey is $\langle z \rangle
\approx 0.15$). The observed H$\alpha$ luminosities are corrected for
dust extinction.   The solid line is the luminosity function from our
model assuming that no ionizing photons escape ($f_{\rm esc,gas}=0$). The
dashed line is the same function for $f_{\rm esc,gas}=0.05$ and the
dotted line for $f_{\rm esc,gas}=0.2$. The H$\alpha$ luminosities from
the model are the values unattenuated by dust.}
\label{fig:HalphaLF}
\end{figure}

\begin{figure}
\psfig{file=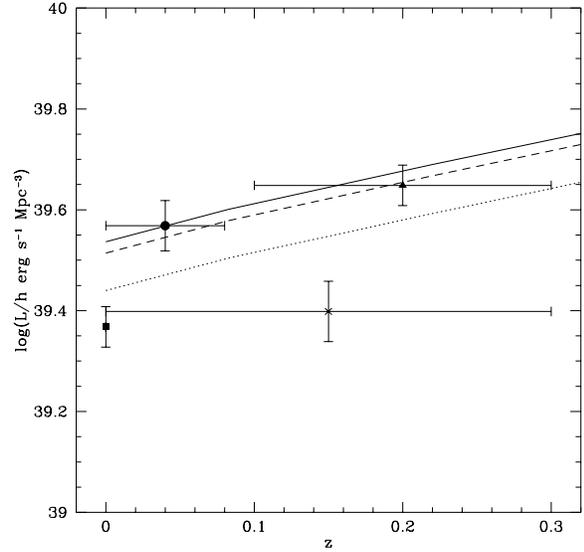,width=80mm}
\caption{The H$\alpha$ luminosity density of the Universe as a
function of redshift. Points with errorbars are observational
estimates, including corrections for
dust extinction: \protect\scite{gallego95} (square);
\protect\scite{tresse98} (triangle);
\protect\scite{gronwall98} (circle) and
\protect\scite{sullivan99} (cross). We have converted the data for
the effects of differing luminosity distances and volume elements to
correspond to the $\Omega_0=0.3$, $\Lambda_0=0.7$ cosmology assumed in
our model.  The solid line is the equivalent luminosity density
measured in our model assuming that no photons escape ($f_{\rm
esc,gas}=0$). The dashed line is the same function for $f_{\rm
esc,gas}=0.05$ and the dotted line for $f_{\rm esc,gas}=0.2$.}
\label{fig:Halphadens}
\end{figure}

In Figs. \ref{fig:HalphaLF} and \ref{fig:Halphadens} we compare the
H$\alpha$ emission line properties of galaxies in our model with
observational data at low redshift. The observed values are already
corrected for dust extinction, so we compare them with the theoretical
values before dust attenuation. In order to calculate these properties
accurately, we simulate halos of mass up to and including
$10^{15}h^{-1}M_\odot$. In calculating the H$\alpha$ line luminosity
of each galaxy, we assume that a fraction $1 - f_{\mathrm esc,gas}$ of
the Lyc photons are absorbed by hydrogen atoms, producing H$\alpha$
photons according to case B recombination. The remaining Lyc photons
escape, after being further attenuated by dust. The figures show
results for $f_{\mathrm esc,gas}=0$, 0.05 and 0.2, which roughly
brackets the likely range of values for typical disk galaxies at the
present day, as we discuss below. Both the predicted H$\alpha$
luminosity function and luminosity density are in reasonable agreement
with the observations, demonstrating that our models produce galaxies
with realistic total ionizing luminosities (before attenuation by gas
and dust). In principle, these observational comparisons provide a
constraint on the value of $f_{\rm esc,gas}$, if the other parameters
in the semi-analytical model are assumed to be known. However, in
practice it is not possible to reliably distinguish between $f_{\rm
esc,gas} = 0.2$ and $f_{\rm esc,gas} = 0$ or $0.05$, given the
uncertainties in the observational data. The observational results
depend on the dust correction factors applied, and there is also some
uncertainty in the ionizing luminosities predicted by stellar
population synthesis models for a given IMF. With these caveats in
mind it would seem that mean escape fractions anywhere between zero
and 20-30\% are acceptable.  Observations of starburst galaxies in the
nearby universe suggest that the escape fraction is actually less than
$3\%$ for such galaxies \cite{leitherer95}, but starbursts are known
to have very high column densities of gas and dust, and so the escape
fraction in normal galaxies can probably be significantly higher
(e.g. \pcite{kenn98}).

\subsection{The dependence of ${\bf f}_{\mathbf esc,gas}$ on redshift and
 on halo mass}
\label{sec:fescmodels}

So far, we have assumed that $f_{\rm esc,gas}$ is a global constant,
varying neither with galaxy properties nor redshift. The details of
the physical processes which determine $f_{\rm esc,gas}$ are
uncertain, but a constant $f_{\rm esc,gas}$ seems unrealistic, as the
properties of the emitting galaxies depend strongly upon both redshift
and the mass of the halos in which they live.  Given the complexity of
this problem, here we merely aim at establishing the general trend of
how $f_{\rm esc,gas}$ may vary with halo mass and redshift.

We will consider three models for $f_{\rm esc,gas}$. In the first
model, $f_{\rm esc,gas}$ is assumed to be a universal constant (this
will be referred to as the ``fixed model''). In the second and third
models $f_{\rm esc,gas}$ is evaluated for each galaxy, based on its
physical properties. These two models are described next.

Our first physical $f_{\rm esc,gas}$ model is based on the approach of
\scite[hereafter DS94]{doveshull94} who derived an analytic
expression for $f_{\rm esc,gas}$. In their model, Lyc photons are
emitted by OB associations in a galactic disk and escape by ionizing
``\HII\ chimneys'' in the \HI\ layer. The fraction of photons escaping
a disk of given size and gas content can then be calculated. Whilst
the original DS94 model assumes that OB associations all lie in the
mid-plane of the galaxy disk, we have also considered the case where
OB associations are distributed vertically like the gas in the disk.

The \HII\ chimney model of DS94 does not include the effects of finite
lifetimes of the OB associations, or of dynamical evolution of the gas
distribution around an OB association due to energy input by stellar
winds and supernova from the OB association itself. \scite{ds99}
have calculated the escape of ionizing photons through a dynamically
evolving superbubble, which is driven by an OB association at its
centre. They find that the resulting escape fractions are slightly
lower than those obtained from the DS94 model (since the superbubble
shell is able to effectively trap radiation). Numerical solutions of
the radiative transfer equations in disk galaxies give results in
excellent agreement with the Str\"omgren sphere approach of DS94 for
OB associations at the bright end of the luminosity function, but give
somewhat lower escape fractions for the faintest OB associations, the
two approaches differing by around 25\% for a single OB star
\cite{wood99}.

Our second physical model for $f_{\rm esc,gas}$ is based on
\scite[hereafter DSGN98]{devriendt99}. In this case, the ionizing
stars are assumed to be uniformly mixed with the gas in the galaxy,
and the gas is assumed to remain neutral. DSGN98 give an approximate
analytic expression for the escape fraction in this case, but we have
instead calculated the escape fraction exactly by numerical
integration, for a specific choice for the gas density profile.

We give details of the calculation of $f_{\rm esc,gas}$ in the DS94
and DSGN98 models in Appendix \ref{append:fesc}. Both models contain
one free parameter, $h_{\rm z}/r_{\rm disk}$, the ratio of disk scale
height to radial scale length. We will consider the effects of varying
this parameter in \S\ref{sec:variants}. For starbursts, we calculate
the escape fraction based on a simple spherical geometry, as is also
described in Appendix \ref{append:fesc}. The contribution to the total
ionizing luminosity from bursts of star formation is small ($<8\%$) at
all redshifts.

\begin{figure}
\psfig{file=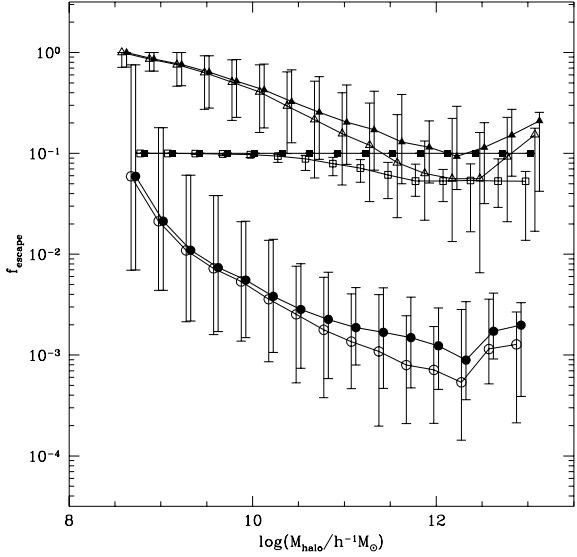,width=80mm}
\caption{The escape fraction, $f_{\mathrm esc}$, at $z=0$ as a
function of halo mass. Thick lines with solid symbols show the escape
fraction ignoring the effects of dust, whilst thin lines with open
symbols include absorption by dust. Three models for the gas escape
fraction are plotted: fixed escape fraction of 10\% (squares), DS94
(triangles), and DSGN98 (circles). In each case, the symbols indicate
the median of the distribution of escape fractions, whilst the
errorbars indicate the 10\% and 90\% intervals of the distribution.}
\label{fig:fescmass}
\end{figure}

To summarize, we will show results from three models for $f_{\rm
esc,gas}$ as standard. These are: a model in which $f_{\rm esc,gas}$
is held constant at 0.1; the DS94 model with OB associations in the
disk midplane; and the DSGN98 model using our exact calculation of the
escaping fraction.  We consider the DS94 model to be the most
realistic of our three models for $f_{\rm esc,gas}$, but also present
results from the other models for comparison.

In Fig. \ref{fig:fescmass} we show the variation of \fesctxt\ with
dark halo mass at $z=0$ for the three models. The thin and thick lines
show the escape fraction respectively with and without attenuation by
dust. When a halo contains more than one galaxy, we plot the mean
\fesctxt\ weighted by ionizing luminosity. At a given halo mass, halos
with the lowest \fesctxt\ tend to have the highest ionizing
luminosities, as both the star formation rate and attenuation of
photons are increased in galaxies with large gas contents. The three
models all show a trend for decreasing escape fraction with increasing
halo mass up to $M_{\rm halo}\sim 10^{12}h^{-1}M_\odot$. For the fixed
gas escape fraction model, the variation in \fesctxt\ is due entirely
to the effects of dust, which can therefore be seen to be negligible
in halos less massive than $\sim 10^{10} h^{-1} M_{\odot}$. This
decrease in \fesctxt\ due to dust is enhanced in the other two models
by the variation in $f_{\rm esc,gas}$, which also declines with
increasing halo mass. For halos more massive than
$10^{12}h^{-1}M_\odot$, the escape fractions rise somewhat for the
variable $f_{\rm esc,gas}$ models. Note that the DSGN98 model predicts
a much smaller escape fraction than the DS94 model at all masses.

\begin{figure}
\psfig{file=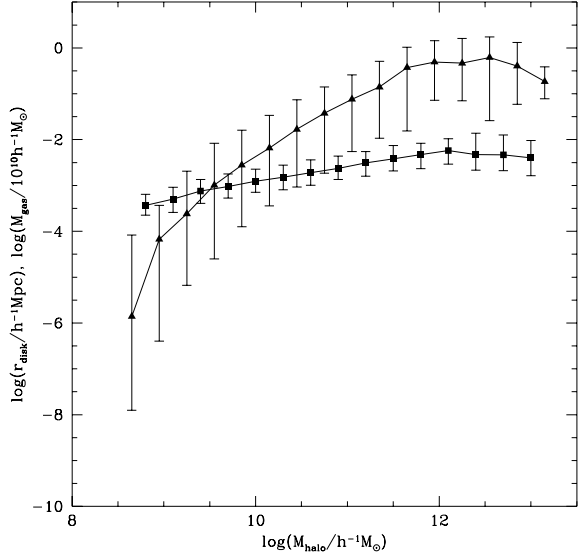,width=80mm}
\caption{The variation
in (ionizing luminosity-weighted) mean disk scale length (squares)
and gas mass (triangles) for galaxies in our model as a function
of halo mass. The symbols show the medians of the distributions,
while the errorbars indicate the 10\% and 90\% intervals.}
\label{fig:fescmasscauses}
\end{figure}

In Fig. \ref{fig:fescmasscauses} we plot the variation in the
(ionizing luminosity-weighted) mean disk scale length and cold gas
mass for galaxies in our model as a function of halo mass. Evidently,
the decline in \fesctxt\ with increasing halo mass below
$10^{12}h^{-1}M_\odot$ seen in Fig.~\ref{fig:fescmass} is due mainly
to the greater masses of gas found in galaxies in these halos. This
rapid change in the mass of gas present is due to the effects of
feedback, which efficiently ejects gas from galaxies in low mass
halos. Above $10^{12}h^{-1}M_\odot$, the mass of cold gas in galaxies
levels off and then begins to decline as cooling becomes inefficient
in more massive halos. This results in an escape fraction increasing
with halo mass for the most massive halos simulated. Although galaxy
sizes increase with increasing halo mass, thereby reducing gas
densities somewhat, this effect is not strong enough to offset the
increased cold gas mass in these galaxies.

\begin{figure}
\psfig{file=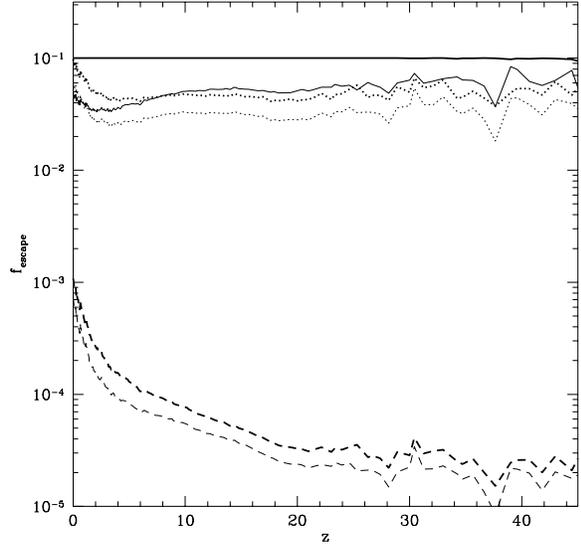,width=80mm}
\caption{The (ionizing luminosity-weighted) mean escape fraction for all galaxies as a function of redshift. Thick lines show the escape fraction ignoring the effects of dust, whilst thin lines include dust. Three models for the gas escape fraction are plotted: constant gas escape fraction of 10\% (solid line), DS94 (dotted line) and DSGN98 (dashed line).}
\label{fig:fescz}
\end{figure}

We find that the DS94 and DSGN98 models applied to our galaxies
predict escape fractions (including the effects of dust) for halos of
mass $\sim 10^{11} h^{-1} M_{\odot}$ at $z=0$ of $\approx 20\%$ and
$\approx 0.2\%$ respectively.  The mean DS94 and DSGN98
luminosity-weighted escape fractions for galaxies at $z=0$ are lower,
being $\approx 6\%$ and $\approx 0.1\%$ respectively. However, we
expect some variation in these values with redshift due to the
evolution of the galaxy population. In fact, we find a rapid decline
in both cold gas content and galaxy disk size with increasing
redshift. In Fig. \ref{fig:fescz} we show the evolution of the
(ionizing luminosity-weighted) mean \fesctxt\ between redshifts 0 and
45. All models show an initial rapid decline in $f_{\rm esc}$ with
increasing $z$. After this, in the constant $f_{\rm esc,gas}$ model,
the mean escape fraction increases with redshift since the dust
content of galaxies was lower in the past. The DS94 model shows a very
gradually rising escape fraction, whilst the DSGN98 model has a more
rapid decline.

In our model, the contribution of stellar sources to the UV background
is dominated by galaxies at low redshifts ($z\lsim 1$). We find that
immediately shortwards of 912\AA\, our DS94 model predicts a
background due to stellar sources which is very close to that expected
from QSOs \cite{hm96}, after including the effects of attenuation by
the intervening IGM \cite{madau95}. At shorter wavelengths the QSO
contribution soon becomes dominant. Thus, at $z=0$ the combined
background due to stars (from our DS94 model) and QSOs
(from \pcite{hm96}) is $J_{\hbox{\scriptsize 912\AA}} \approx
4\times 10^{-23}$ ergs/s/cm$^2$/Hz/ster. This is consistent with the
upper limit of $J_{\hbox{\scriptsize 912\AA}} = 8\times 10^{-23}$
ergs/s/cm$^2$/Hz/ster found by \scite{vogel95}, who searched for
H$\alpha$ emission from intergalactic HI clouds. The contribution of
galaxies to the local ionizing background has also been estimated by
\scite{giallongo97}, based on the luminosity function of galaxies
observed in the Canada-France Redshift Survey. They estimated the
galactic contribution as $J_{\hbox{\scriptsize 912\AA}} \approx5\times
10^{-23}$ ergs/s/cm$^2$/Hz/ster at $z=0$, assuming an escape fraction
of $f_{\rm esc}=0.15$. If we assume the same $f_{\rm esc}$ in our
model, we obtain $J_{\hbox{\scriptsize 912\AA}}=5.2\times 10^{-23}$
ergs/s/cm$^2$/Hz/ster, in excellent agreement with their result.

\section{The filling factor and the evolution of the ionization fronts}
\label{sec:fronts}

We define the filling factor, $F_{\rm fill}$, as the fraction of
hydrogen in the IGM (as defined in \S\ref{sec:IGM}) which has been
ionized. This is the natural quantity which serves as an indication of
the amount of reionization in the IGM. We calculate the growth of the
ionized region around each halo, using the ionizing luminosities
predicted by the semi-analytic model, and then sum over all halos to
find $F_{\rm fill}$. We make two simplifying assumptions: (1) the
radiation from each halo is emitted isotropically, and (2) the
distribution of hydrogen is uniform on the scale of the ionization
front and larger (but with small-scale clumping). It follows that each
halo by itself would produce a spherical ionization front.

The mass of hydrogen ionized within the ionization front, $M$, in
spherical symmetry is given by \cite{shapiro87,haiman96}
\begin{equation}
{1 \over {\rm m}_{\rm H}} {{\rm d}M \over {\rm d}t} = S(t) - \alpha_{\rm H}^{(2)} a^{-3} f_{\rm clump} n_{\rm H} {M \over {\rm m}_{\rm H}},
\label{ion_cgs}
\end{equation}
where $n_{\rm H}$ is the comoving mean number density of hydrogen
atoms (total, \HI\ and \HII) in the IGM, $a(t)$ is the scale factor of
the universe normalized to unity at $z=0$, $t$ is time and $S(t)$ is
the rate at which ionizing photons are being emitted. The factor
$f_{\rm clump}(t)$, defined by
\begin{equation}
f_{\rm clump} = \langle \rho_{\rm IGM}^2\rangle/\bar{\rho}_{\rm IGM}^2,
\label{eq:clumpdef}
\end{equation}
is the clumping factor for the ionized gas in the IGM (here $\rho_{\rm
IGM}$ is the density of IGM gas at any point and $\bar{\rho}_{\rm
IGM}$ is the mean density of the IGM). Small-scale clumpiness causes
the total recombination rate to be larger than for a uniform medium of
the same mean density.

The value of $f_{\mathrm clump}(t)$ for the ionized gas in the IGM is
complicated to calculate analytically. We remind the reader that in
our picture, the IGM consists of all gas which has \emph{not} been
collisionally ionized in halos nor become part of a galaxy. For a
uniform IGM $f_{\mathrm clump}=1$ by definition. If low density
regions of the IGM are ionized before high density regions, as
suggested by \scite{miralda00}, then this would be similar to
having $f_{\mathrm clump}<1$ in eqn.~(\ref{ion_cgs}), but for most
purposes, $f_{\mathrm clump}=1$ can be considered as an approximate
lower bound. We make two different estimates of the possible effects
of clumping.

For our first estimate, which we call $f_{\rm clump}^{\rm
(variance)}$, we assume that the photoionized gas basically traces the
dark matter, except that gas pressure prevents it from falling into
dark matter halos with virial temperatures smaller than $10^4$K (the
approximate temperature of the photo-ionized gas). Thus, we calculate
the clumping factor as $f_{\rm clump}^{\rm (variance)} = (1 +
\sigma^2)$, where $\sigma^2$ is the variance of the dark matter
density field in spheres of radius equal to the virial radius of a
$10^4$K halo. $\sigma^2$ is calculated from the non-linear dark matter
power spectrum, estimated using the procedure of \scite{pnd96},
and smoothed using a top-hat filter in real space.

For our second estimate, which we call $f_{\rm clump}^{(\rm halos)}$,
we include the effects of collisional ionization in halos and of
removal of gas by cooling into galaxies in a way consistent with our
definition of $f_{\rm IGM}$ given in equation~(\ref{eq:fIGM}). The
diffuse gas in halos with virial temperatures above $\approx 10^4$K is
assumed to have the density profile of an isothermal sphere with a
constant density core. The gas originally associated with smaller
halos is assumed to be pushed out of these halos by gas pressure
following photoionization, and to be in a uniform density component
occupying the remaining volume. As shown in
Appendix~\ref{append:fclump}, the clumping factor is then
\begin{eqnarray}
f_{\rm clump}^{(\rm halos)} & = & {f_{\rm m,smooth}^2 \over f_{\rm v,smooth} f_{\rm IGM}^2} + {f_{\rm int} \Delta_{\rm vir} \over f_{\rm IGM}^2} \int_{M_{\rm J}}^\infty \langle (1-f_{\rm gal})^2 \rangle \nonumber \\
 & & \times (1-x_{\rm H})^2 {M_{\rm halo} \over \rho_{\rm c} \Omega_0} {{\rm d}n \over {\rm d}M_{\rm halo}} {\rm d}M_{\rm halo},
\end{eqnarray}
where $M_{\rm J}$ is the mass of a halo which just retains reionized
gas, $f_{\rm m,smooth}$ is the fraction of the total baryonic mass in
the uniform component, and $f_{\rm v,smooth}$ is the fraction of the
volume of the universe occupied by this gas. Here, $f_{\rm gal}$ is
the fraction of the baryonic mass in a halo in the form of galaxies,
$x_{\rm H}$ is the fraction of hydrogen in the diffuse halo gas which
is collisionally ionized (as in eqn.~\ref{eq:fIGM}), and $\langle
\rangle$ indicates an average over all halos of mass $M_{\rm
halo}$. The factor $f_{\rm int}$ is a parameter depending only on the
ratio of the size of the core in the gas density profile to the halo
virial radius. For a core radius equal to one-tenth of the virial
radius $f_{\rm int}=3.14$ (see Appendix~\ref{sec:clumphalo}). This
estimate of the clumping factor ignores the possibility of gas in the
centres of halos (but not part of a galaxy) becoming self-shielded
from the ionizing radiation.  Such gas would not become photoionized,
and so would not contribute to the recombination rate, resulting in
$f_{\rm clump}$ being lower than estimated here. A detailed treatment
of the ionization and temperature structure of gas inside halos is
beyond the scope of this work.

Of course, before reionization, when the gas is typically much cooler
than $10^4$K, gas will fall into dark matter halos with virial
temperatures below $10^4$K. If all gas were in halos, then we would
find
\begin{eqnarray}
f_{\rm clump} & \approx & {f_{\rm int} \Delta_{\rm vir} \over f_{\rm IGM}^2} \int_0^\infty \langle (1-f_{\rm gal})^2 \rangle (1-x_{\rm H})^2 {M_{\rm halo} \over \rho_{\rm c} \Omega_0} \nonumber \\
 & & \times {{\rm d}n \over {\rm d}M_{\rm halo}} {\rm d}M_{\rm halo}
\end{eqnarray}
Once reionized, some of the gas in these small halos will flow back
out of the halo as the gravitational potential is no longer deep
enough to confine the gas.

We consider $f_{\rm clump}^{(\rm halos)}$ as our best estimate of the
IGM clumping factor, at least for lower redshifts, when a significant
fraction of the gas is in halos with $M>M_J$, while $f_{\rm
clump}^{\rm (variance)}$ is more in the nature of an upper limit.

\begin{figure}
\psfig{file=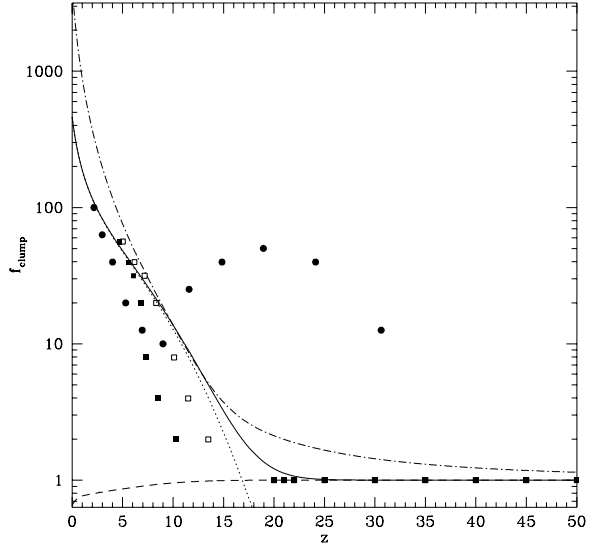,width=80mm}
\caption{The gas clumping factor $f_{\mathrm clump}$ as a function of
redshift. The solid line shows $f_{\rm clump}^{\rm (halos)}$, whilst
the dotted and dashed lines show the contributions to this quantity
from gas inside and outside halos respectively. The dot-dashed line
shows $f_{\rm clump}^{\rm (variance)}$. Filled circles show the
clumping factor calculated by \protect\scite{valageas99}. Squares
show the clumping factors determined from a simulation by
\protect\scite{gnedin97} for all baryons (open squares) and
baryons in ionized regions only (filled squares).}
\label{fig:fclump}
\end{figure}

The clumping factors $f_{\rm clump}^{(\rm halos)}$ and $f_{\rm
clump}^{\rm (variance)}$ are plotted as functions of redshift in
Fig.~\ref{fig:fclump}. They show fairly similar behaviour above
$z\approx 10$. Below this redshift, $f_{\rm clump}^{\rm (variance)}$
greatly exceeds $f_{\rm clump}^{(\rm halos)}$, because it becomes
dominated by gas in massive dark matter halos, which, on the other
hand, contributes negligibly to $f_{\rm clump}^{(\rm halos)}$ as it is
collisionally ionized. We also plot estimates of the clumping factor
from two other papers: \scite{valageas99} calculated the clumping
factor of the baryons which have been unable to cool (the quantity
they call $C_n$) using their own analytical model. They obtain values
of $f_{\rm clump}$ which are comparable to $f_{\rm clump}^{\rm
(halos)}$ at $z\lsim 10$, but are substantially larger at higher
redshifts. \scite{gnedin97} performed hydrodynamical simulations
in a cosmology similar to that which we consider, from which they
measured $f_{\rm clump}$ directly. They calculated two clumping
factors: one for all baryons in their simulation, $f_{\rm
clump}=f_{\rm clump}^{\rm (GO:bb)}$, and the other for baryons in
ionized regions only, $f_{\rm clump}=f_{\rm clump}^{\rm (GO:H_{\sc
II})}$, which is smaller.  $f_{\rm clump}^{\rm (GO:H_{\sc II})}$ is
more relevant for our purposes, but may still overestimate the
clumping of photo-ionized gas in the IGM, since it includes
collisionally ionized gas in galaxy halos. These clumping factors are
everywhere lower than $f_{\rm clump}=f_{\rm clump}^{\rm (variance)}$.
$f_{\rm clump}^{\rm(GO:H_{\sc II})}$ is close to our estimate $f_{\rm
clump}^{(\rm halos)}$ at the highest and lowest redshifts, but smaller
in the intermediate range.

\subsection{Model results}
\label{sec:ffresults}

\begin{figure}
\psfig{file=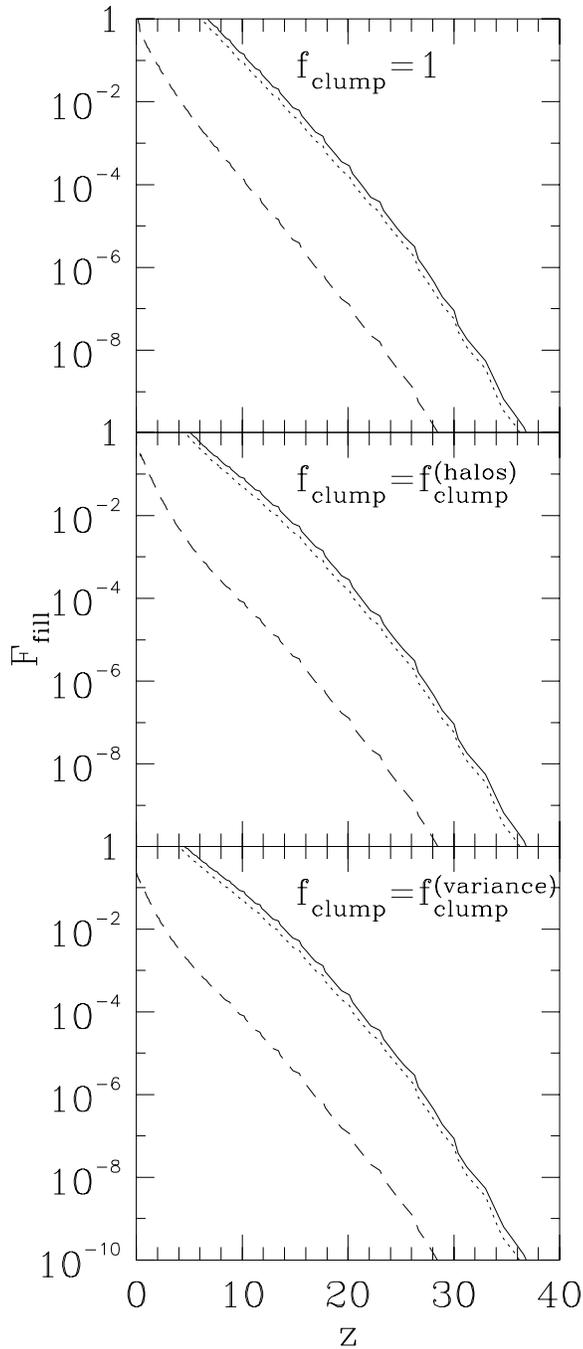,width=80mm,bbllx=0mm,bblly=10mm,bburx=110mm,bbury=270mm,clip=}
\caption{The filling factor, $F_{\rm fill}$, as a function of redshift
determined using three different models for the absorption of ionizing
photons by gas inside galaxies: constant gas escape fraction of 10\%
(solid lines), DS94 (dotted lines) and DSGN98 (dashed lines). All
models include the effects of dust on the escape fraction. The three
panels show the filling factors for three different assumptions about
the IGM clumping factor, $f_{\rm clump}$: no clumping (top panel),
clumping due to virialised halos (middle panel) and clumping estimated
from the \protect\scite{pnd96} non-linear power spectrum (bottom
panel).} 
\label{fig:Ffill}
\end{figure}

In Fig. \ref{fig:Ffill} we show the ionized filling factor of the IGM,
$F_{\rm fill}$, as a function of redshift. Here we compute $F_{\rm
fill}$ by summing the volumes of the \HII\ regions formed around each
halo, weighted by the number of such halos per unit volume as given by
the Press-Schechter theory. (Later we will use the halo mass function
measured directly from an N-body simulation to calculate $F_{\rm
fill}$ --- see Fig.~\ref{fig:ffillsim}). $F_{\rm fill}$ will exceed 1
if more ionizing photons have been produced than are needed to
completely reionize the universe. We show results for our three models
for $f_{\rm esc,gas}$, and for three different assumptions about
$f_{\rm clump}$.

If we ignore the effects of dust, we find that the model with a
constant escape fraction of 10\% reionizes the Universe by $z=7.9$ if
$f_{\rm clump}=1$ but only by $z=6.6$ if $f_{\rm clump}=f_{\rm
clump}^{\rm (halos)}$. In order to reionize the Universe by $z=5$,
escape fractions of 1.4\% and 3.7\% are needed for $f_{\rm clump}=1$
and $f_{\rm clump}=f_{\rm clump}^{\rm (halos)}$ respectively. When we
include the effects of dust, we find that gas escape fractions of
3.3\% and 9.3\% are needed to reionize by $z=5$ for these two cases.

If, instead of assuming a constant gas escape fraction, we use the
more physically motivated DS94 model, we find reionization occurs at
$z=6.1$ if $f_{\rm clump}=1$, but only at $z=4.5$ if $f_{\rm
clump}=f_{\rm clump}^{\rm (halos)}$ (both estimates including dust).
In the latter case, the ionized filling factor at $z=5$ is only
76\%. If we assume OB associations are distributed as the gas in the
DS94 model (as opposed to lying in the disk mid-plane as in our
standard model) then a filling factor of 117\% (i.e. full
reionization) is achieved by $z=5$. The DSGN98 model, which predicts
much lower escape fractions than the DS94 model at all redshifts, is
able to reionize only $\approx 2$\% of the IGM by redshift 3, and full
reionization never occurs even if $f_{\rm clump}=1$.

We note that in the DS94 model, approximately 90\% of the photons
required for ionization are produced at $z<10$. Thus our neglect of
radiative feedback effects (which may reduce the number of ionizing
photons produced at higher redshifts) is unlikely to seriously effect
our determination of the reionization epoch.

As both the DS94 and DSGN98 models predict quite low escape fractions,
we have also considered a much more extreme model which simply assumes
that $f_{\rm esc}=\beta/(1+\beta)$, where $\beta$ is the feedback
efficiency as defined by eqn.~(\ref{eq:beta}). This toy model, which
we will refer to as the ``holes scenario'', produces very high
escaping fractions for galaxies with low circular speeds, and low
escaping fractions for those with high circular speeds. A behaviour
for $f_{\rm esc}$ of this general form might result if photons are
able to escape through holes in the galaxy disk which have been
created by supernovae. Since dust would also be expected to be swept
out of these holes we do not include any dust absorption in this
model. The holes scenario produces very different results compared to
our two physical models for the escape fraction. In this model $f_{\rm
esc}\approx 1$ for $z>10$, dropping to 45\% by $z=0$. Not surprisingly
therefore, this model succeeds in satisfying the Gunn-Peterson
constraint, reionizing the Universe by $z=11.7$ if $f_{\rm clump}=1$
and by $z=10.6$ if $f_{\rm clump}=f_{\rm clump}^{\rm (halos)}$. While
this model is only a very crude attempt to consider a dynamically
disturbed gas distribution in galaxy disks, it clearly demonstrates
that such effects may be of great importance for studies of
reionization.

We have also computed the filling factor in our model using the
clumping factors calculated by \scite{gnedin97} and
\scite{valageas99} (as given in Fig. \ref{fig:fclump}). Of
course, this is not strictly self-consistent, as their clumping
factors are calculated from their own models for galaxy formation and
reionization, which differ from ours. Using either of these with the
DS94 model gives a reionization redshift comparable to that obtained
using $f_{\rm clump}=f_{\rm clump}^{\rm (halos)}$: we find
reionization at $z=3.6$ using the \scite{gnedin97} clumping
factor, and $z=4.9$ using that of \scite{valageas99}.

\begin{figure}
\psfig{file=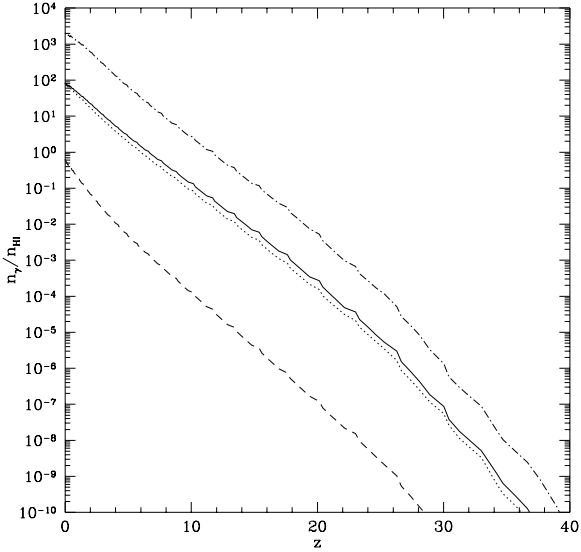,width=80mm}
\caption{The ratio of the total number of ionizing photons which have escaped into the IGM per
comoving volume by redshift $z$ to the comoving number density of
hydrogen nuclei in the IGM. The dot-dash line assumes that all ionizing
photons can escape from their galaxies into the IGM. Three models for
the gas escape fraction are also shown (each also includes the effects
of dust): fixed escape fraction of 10\% (solid line), DS94 (dotted
line) and DSGN98 (dashed line).}
\label{fig:ngnH}
\end{figure}

In Fig. \ref{fig:ngnH} we show the total number of ionizing photons
which have escaped into the IGM per unit comoving volume by redshift
$z$, $n_\gamma$, divided by the total number of hydrogen nuclei in the
IGM per unit comoving volume, $n_{\rm H}$ (which is $f_{\rm IGM}$
times the total number density of hydrogen nuclei). When this number
reaches one, just enough photons have been emitted by galaxies to
reionize the IGM completely if recombinations are unimportant. This
criterion has been used previously to estimate when reionization may
occur. Since our model includes the effects of recombinations in the
IGM, we can judge how well this simpler criterion performs. If we
ignore the effects of absorption by gas and dust on the number of
ionizing photons escaping from galaxies, we find that, in this
cosmology, our model achieves $n_{\gamma}/n_{\rm H} = 1$ by $z\approx
12$. When we account for the effects of dust and gas in galaxies, we
find that the redshift at which $n_{\gamma}/n_{\rm H} = 1$ is
significantly reduced, the exact value depending on the model for the
escape fraction.  With a fixed gas escape fraction of $10\%$,
$n_{\gamma}/n_{\rm H} = 1$ by $z
\approx 7$, whilst for the DS94 model $n_{\gamma}/n_{\rm H} = 1$ is
achieved at $z\approx 6$. In the DSGN98 model $n_{\gamma}/n_{\rm H} =
1$ has not been achieved even by $z=0$. When we include recombinations
in the IGM, the model with constant gas escape fraction reaches
$F_{\rm fill} = 1$ only by $z=6.7$ for $f_{\rm clump}=1$, and by
$z =5.1$ for $f_{\rm clump}=f_{\rm clump}^{\rm (halos)}$, as shown in
Fig. \ref{fig:Ffill}, showing how reionization is delayed.

Note that while the cosmology considered here is similar to Model G of
\scite{baugh98}, the parameters of the semi-analytic model used
are somewhat different. Specifically, \scite{baugh98} used a
model in which feedback was much more effective in low mass halos than
in our model, since they required their models to produce a B-band
luminosity function with a shallow faint end slope. As a result, the
epoch at which $n_{\gamma}/n_{\rm H} = 1$ was much later in Model G of
\scite{baugh98} than in our current model.

In summary, we see that, even for a specific model of galaxy
formation, the predicted epoch of reionization is sensitive to the
uncertain values of the escape fraction \fesctxt\ and the clumping
factor $f_{\rm clump}$. If the clumping factor is as large as $f_{\rm
clump}^{\rm (halos)}$, then in the case of a constant gas escape
fraction $f_{\rm esc,gas}$, we need $f_{\rm esc,gas}\gsim 10\%$ in our
model to ionize the IGM by $z=5$, if absorption by dust is included,
and $f_{\rm esc,gas}\gsim 4\%$ if dust is ignored. With the more
physically-motivated DS94 and DSGN98 models, and the same clumping
factor, at most 76\% of the IGM is reionized by $z=5$, which would be
inconsistent with observations of the Gunn-Peterson effect. For the
extreme case of a uniform IGM, reionization occurs by $z=6.1$ even
with the DS94 model for \fesctxt.  Our ``best estimate'' is based on
combining the DS94 model with $f_{\rm clump}^{\rm (halos)}$ for the
IGM clumping factor. As already stated, this model narrowly fails to
satisfy the Gunn-Peterson constraint at $z=5$ (unless we assume that
OB associations are distributed as the gas, rather than lying in the
disk mid-plane), suggesting that additional sources of ionizing
radiation are required at high redshift, either more stars than in our
standard model, or non-stellar sources (e.g. quasars). However, given
the theoretical uncertainties in \fesctxt\ and $f_{\rm clump}$, we
consider that this is not yet proven.

\section{Sensitivity of results to model parameters}
\label{sec:variants}

We turn now to test the robustness of our results to variations in the
parameters of our galaxy formation model.  To do this, we have varied
key parameters of the models and determined the ionized hydrogen
filling factor in each case. We consider several different models. The
variant models which we consider are listed in Table
\ref{tb:varmods}. In each case, we give the value of the parameter
which is changed relative to the standard model given in Table
\ref{tb:standpars}.

\begin{figure}
\psfig{file=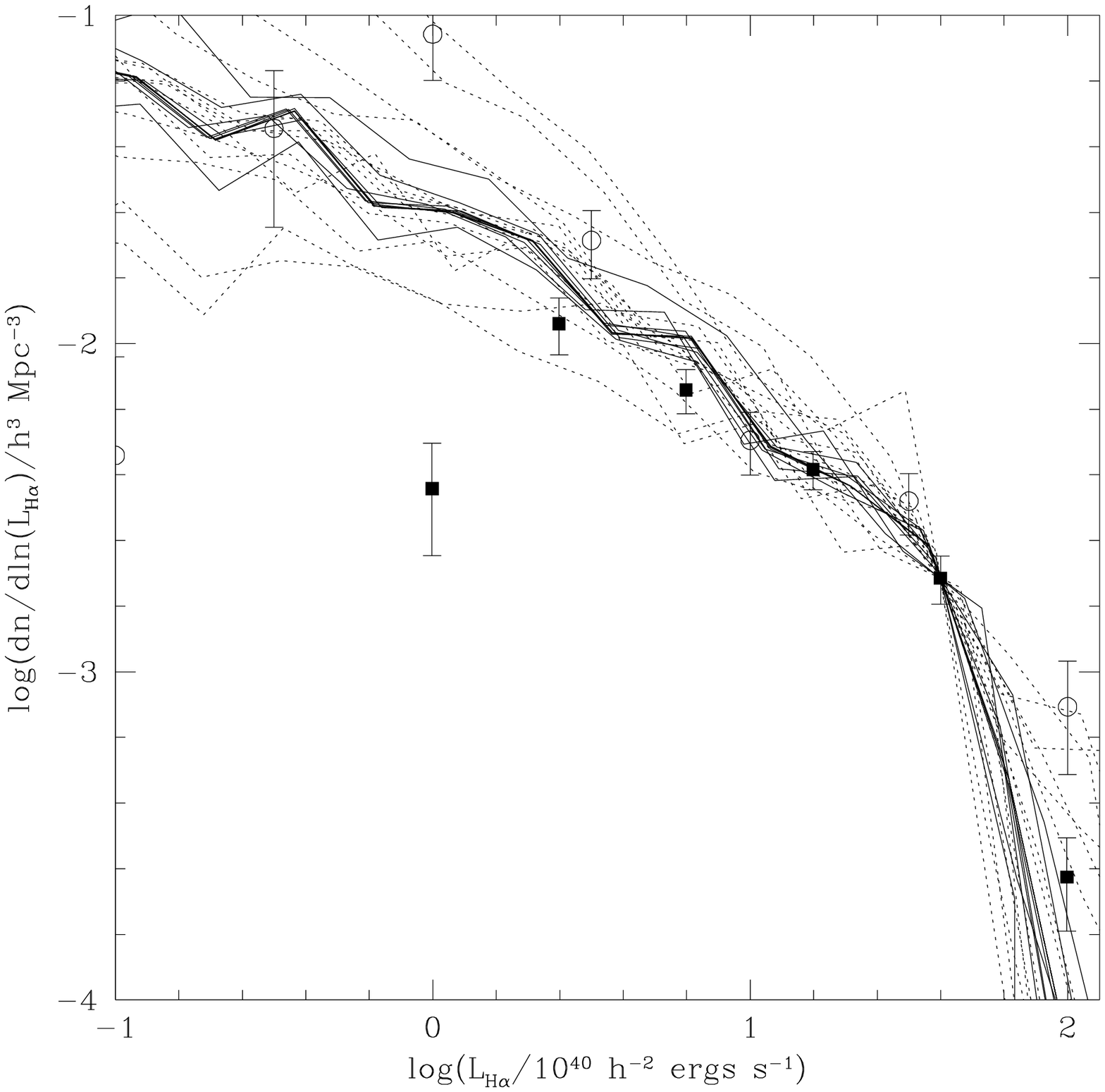,width=80mm}
\caption{The $z=0$ H$\alpha$ luminosity functions of our variant
models. The solid lines show the variant models, whilst the dotted
lines indicate the variants models marked with a $\dag$ in Table
\protect\ref{tb:varmods}. Points with error bars are observational
data from \protect\scite{gallego95} (filled squares) and
\protect\scite{sullivan99} (open circles). These points include a
correction for dust extinction.}
\label{fig:HaLFvar} 
\end{figure}

\scite{ajbclust} have shown that normalising models to the $z=0$
B-band luminosity function allows robust estimates of the $z=0$ galaxy
correlation function to be made. Here we choose a similar constraint,
forcing all models to match the $z=0$ H$\alpha$ luminosity function of
\scite{gallego95} at $L_{\rm H\alpha} = 4 \times 10^{41} h^{-2}$
ergs/s (note that at these luminosities the Gallego et al. luminosity
function agrees, within the errorbars, with that of
\pcite{sullivan99}). This is achieved by adjusting the value of the
parameter $\Upsilon$ (which determines the fraction of brown dwarfs
formed in the model). The $z=0$ H$\alpha$ luminosity functions for all
models considered are shown in Fig. \ref{fig:HaLFvar}. Dotted lines
show those models with H$\alpha$ luminosity functions that are
significantly different from that of the standard model (at either the
bright or faint ends).

In Table \ref{tb:varmods} we list escape fractions and filling factors
in the variant models for the fixed and DS94 models for \fesctxt\, for
the case $f_{\rm clump}=f_{\rm clump}^{\rm (halos)}$. Values of
$F_{\rm fill}$ may exceed unity, as, in some models, by $z=5$ more
ionizing photons have escaped into the IGM than are required to
reionize the universe.

\begin{table*}
\caption{Results of the variant models at $z=5$. Models marked by a
\dag\ are those which have very different H$\alpha$ luminosity
functions compared to the standard model (see
Fig. \protect\ref{fig:HaLFvar}). The second column lists those
parameters which differ from the standard model. The third column
lists the value of $\Upsilon$ required for each model to match the
H$\alpha$ luminosity function. Columns 4--5 list the mean escape
fraction (including the effects of dust) in the fixed and DS94
models. Finally, columns 6--7 list the filling factors at $z=0$ in the
fixed and DS94 models, for the case $f_{\rm clump}=f_{\rm clump}^{\rm
(halos)}$. Note that filling factors will exceed 1 if more photons
than required to reionize the universe have been produced.}
\label{tb:varmods}
\begin{center}
\begin{tabular}{clccccc}
\hline
& & & \multicolumn{2}{c}{${\bf f_{\bf escape} (\%)}$} & \multicolumn{2}{c}{${\bf F}_{\bf fill}$} \\
\textbf{Model} & \textbf{Parameter change(s)} & ${\bf \Upsilon}$ & \textbf{Fixed ($\mathbf{f_{\bf esc,gas}=0.1}$)} & \textbf{DS94} & \textbf{Fixed} & \textbf{DS94} \\
\hline
Standard & None & 1.53 & 3.9 & 2.7 & 1.08 & 0.76 \\
1$^\dag$ & $\Omega_{\rm b} = 0.04$ & 2.87 & 1.7 & 0.8 & 0.29 & 0.13 \\
2$^\dag$ & $\Omega_{\rm b} = 0.01$ & 0.36 & 6.4 & 7.6 & 6.49 & 7.60 \\
3$^\dag$ & $\alpha_{\rm hot} = 0.5$ & 1.22 & 2.8 & 1.6 & 1.54 & 0.87 \\
4$^\dag$ & $\alpha_{\rm hot} = 4.0$ & 1.66 & 3.2 & 2.0 & 0.49 & 0.32 \\
5$^\dag$ & $V_{\rm hot} = 300$ km/s & 0.95 & 5.5 & 4.6 & 1.22 & 1.03 \\
6$^\dag$ & $V_{\rm hot} = 50$ km/s & 0.91 & 2.2 & 1.3 & 2.01 & 1.13 \\
7$^\dag$ & $\epsilon_\star = 0.020$ & 1.74 & 3.9 & 3.3 & 1.25 & 1.07 \\
8 & $\epsilon_\star = 0.005$ & 1.19 & 3.9 & 2.4 & 1.02 & 0.60 \\
9 & $f_{\rm df} = 5.0$ & 1.47 & 4.0 & 2.9 & 1.14 & 0.83 \\
10 & $f_{\rm df} = 0.2$ & 1.49 & 2.6 & 1.7 & 1.04 & 0.68 \\
11$^\dag$ & $f_{\rm ellip} = 0.05$ & 0.86 & 2.8 & 2.0 & 1.78 & 1.27 \\
12 & $f_{\rm ellip} = 0.60$ & 1.63 & 4.0 & 2.9 & 1.02 & 0.72 \\
13 & $f_{\rm dyn} = 5.0$ & 1.53 & 4.0 & 2.8 & 1.08 & 0.77 \\
14 & $f_{\rm dyn} = 0.1$ & 1.53 & 4.0 & 2.9 & 1.08 & 0.77 \\
15$^\dag$ & $p = 0.04$ & 1.20 & 3.0 & 2.4 & 1.01 & 0.78 \\
16$^\dag$ & $p = 0.01$ & 1.46 & 5.1 & 3.3 & 1.47 & 0.96 \\
17 & $R = 0.50$ & 1.86 & 3.9 & 2.7 & 0.89 & 0.63 \\
18$^\dag$ & $R = 0.10$ & 1.23 & 4.1 & 3.0 & 1.34 & 0.96 \\
19 & $h_{\rm z} = 0.5$ & 1.53 & 4.0 & 1.6 & 1.08 & 0.42 \\
20$^\dag$ & Feedback: modified $(a_{\rm r}=0.75)$ & 0.58 & 2.7 & 1.5 & 3.25 & 1.82 \\
21 & $S_2 = 20.0\times 10^{50}$ photons/s & 1.53 & 4.0 & 3.4 & 1.08 & 0.91 \\
22 & $S_2 = 0.5\times 10^{50}$ photons/s & 1.53 & 4.0 & 1.4 & 1.08 & 0.36 \\
23 & Starbursts: Not included & 1.60 & 4.0 & 2.9 & 1.04 & 0.73 \\
24 & IMF: \scite{salpeter55} & 1.45 & 4.0 & 2.8 & 1.06 & 0.75 \\
25$^\dag$ & Gas profile: SIS & 1.59 & 2.7 & 1.7 & 1.00 & 0.64 \\
26$^\dag$ & Recooling: allowed & 1.27 & 3.3 & 2.2 & 1.50 & 0.97 \\
\hline
\end{tabular}
\end{center}
\end{table*}

The standard choice for the feedback efficiency, $\beta$, makes
feedback highly efficient in galaxies with low circular velocities. In
this model $\beta = f_{\rm V}^{-\alpha_{\rm hot}}$, where $f_{\mathrm
V} = V_{\mathrm disk} / V_{\mathrm hot}$. The fraction of cold gas
which is reheated by supernovae after infinite time (a quantity with
direct physical interpretation) is then
\begin{equation}
{\beta \over 1-R+\beta} = {f_{\rm V}^{-\alpha_{\rm hot}} \over 1 - R + f_{\rm V}^{-\alpha_{\rm hot}}}.
\end{equation}
Thus as $V_{\rm disk} \rightarrow 0$ all gas is reheated and no stars
are formed. For the modified feedback model, we adapt this form such
that even in arbitrarily small potential wells not all the gas is
reheated by supernovae. We choose
\begin{equation}
{\beta \over 1-R+\beta} = {a_{\rm r} f_{\rm V}^{-\alpha_{\rm hot}} \over 1 - R + f_{\rm V}^{-\alpha_{\rm hot}}},
\label{eq:modfb}
\end{equation}
where $a_{\mathrm r}$ is an adjustable parameter. Now, as $V_{\rm
disk}\rightarrow 0$ a fraction $a_{\rm r}$ of gas is reheated, whilst
a fraction $1-a_{\rm r}$ forms stars. The standard feedback model is
recovered when $a_{\rm r}=1$. For the alternative feedback models
considered here a value of $a_{\rm r}=0.75$ is used.

It should be noted that the variation having one of the greatest
influences on the predicted filling factors is that of Model 20, where
we use the alternative form for feedback given above. This model
produces an H$\alpha$ luminosity function with a very steep faint end
slope, since feedback never becomes highly efficient, even in
extremely small dark matter halos. More ionizing photons are produced
than in the standard model and higher filling factors are achieved.

Other models which alter the strength of feedback (i.e. Models 3, 4, 5
and 6) also cause large changes in the filling factors.  Models with
weaker feedback (i.e. Models 3 and 6) result in larger filling factors
as they allow more star formation to occur in low mass galaxies (these
models again producing a steep slope for the faint end of the
H$\alpha$ luminosity function). The value of $\Omega_{\rm b}$ also has
a strong influence on the filling factors as demonstrated by Models 1
and 2. Finally, in Models 21 and 22 we consider two alternative values
of $S_2$ in the DS94 model. These values span the range of uncertainty
for the maximum OB association luminosity in our own Galaxy
\cite{doveshull94}. These models demonstrate that the filling factors
predicted by the DS94 model are uncertain by a factor of at least 2
simply because of this uncertainty in the value of $S_2$. There is, in
fact, further uncertainty introduced as it is not clear if $S_2$
represents a real cutoff in the  luminosity function of OB associations,
or merely a turn-over in that function.

All of the models which significantly alter the predicted filling
factors are amongst those marked with a $\dag$ in Table
\ref{tb:varmods}, indicating that such models do not reproduce well the
$z=0$ H$\alpha$ luminosity function, and can therefore be discarded as
being unrealistic. With these models removed, our predictions for
$F_{\rm fill}$ are reasonably robust. Considering all the realistic
models we find that for the fixed gas escape fraction of 10\%, $F_{\rm
fill}$ at $z=5$ is $1.08^{+0.06}_{-0.19}$ (where the value indicates
the filling factor in the standard model and the errors show the range
found in the realistic variant models). For the DS94 model we find
$F_{\rm fill}=0.76^{+0.15}_{-0.40}$ (leaving out the models which vary
$S_2$ we find $F_{\rm fill}=0.76^{+0.07}_{-0.34}$).  Our conclusion
that with the DS94 escape fractions reionization cannot happen by
$z=5$ if the clumping factor is as large as $f_{\rm clump}^{\rm
(halos)}$ remains valid under all realistic parameter variations
considered here. On the other hand, if the clumping factor is closer
to the case of a uniform IGM, then reionization by $z=5$ is possible
in the DS94 model (but not in the DSGN98 model).

So far we have considered a single cosmology, namely
$\Lambda$CDM. This choice was motivated by the work of
\scite{coleetal99} and \scite{ajbclust}, who have shown that
the semi-analytic model is able to reproduce many features of the
observed galaxy population for this cosmology. However, in order to
explore the effects of cosmological parameters on reionization, we
have also considered a $\tau$CDM cosmology, with $\Omega=1$, in which
we use model parameters identical to those of
\scite{ajbclust}. We note that \scite{ajbclust} were unable
to match the galaxy correlation function at $z=0$ for this cosmology.

We find that in the $\tau$CDM model, our basic results are unchanged,
i.e. with physical models for the escape fraction, the IGM is
reionized by $z=5$ only if it is much less clumped than in our halo
clumping model with $f_{\rm clump}=f_{\rm clump}^{\rm (halos)}$.  The
escape fractions in this cosmology are actually somewhat higher than
in the $\Lambda$CDM cosmology (due to a lower amount of gas and dust
in galaxies). At $z=5$ the DS94 model predicts a mean escape fraction
$\approx 16\%$, whilst the DSGN98 model predicts $\approx 0.1\%$.
However, the filling factors are significantly lower (for example, in
the DS94 model $F_{\rm fill}=0.24$ at $z=5$ when $f_{\rm clump}^{\rm
(halos)}$ is used as compared to $F_{\rm fill}=0.76$ in
$\Lambda$CDM). This reflects the fact that many fewer ionizing photons
are produced in this cosmology (due to the fact that a stronger
feedback is required in order that the model fits the properties of
galaxies at $z=0$), and that less of the gas has become collisionally
ionized in virialised halos in $\tau$CDM than in $\Lambda$CDM. The
only factor which works in favour of a higher filling factor in
$\tau$CDM is that the clumping factor is somewhat lower. However, this
is not enough to offset the two effects described above.

\section{Spatial distribution of ionizing sources and CMB fluctuations}
\label{sec:spatial}

\subsection{Spatial distribution}
\label{sec:spatdist}

We now consider the temperature anisotropies imprinted on the
microwave background by the IGM following reionization. These depend
on the spatial and velocity correlations of the ionized gas.  A fully
self-consistent calculation of these correlations on the relevant
scales would require very high resolution numerical simulations
including both gas dynamics and radiative transfer
(e.g. \pcite{abel99}). No such numerical simulation is yet
available with the necessary combination of volume and resolution to
calculate the secondary CMB anisotropies on all angular scales of
interest. Therefore in this paper, we calculate the spatial and
velocity distribution of the ionized gas in an approximate way, by
combining our semi-analytical galaxy formation model with a high
resolution N-body simulation of the dark matter.

We have used the same $\Lambda$CDM simulation as \scite[described
in detail by Jenkins et al. 1998]{ajbclust}, which has $\Omega _0 =
0.3$, a cosmological constant $\Lambda _0 = 0.7$, a Hubble constant of
$h=0.7$ in units of $100{\mathrm km/s/Mpc}$, and which is normalised
to produce the observed abundance of rich clusters at $z\approx 0$
\cite{eke96}. Using the same semi-analytic model as employed here,
\scite{ajbclust} were able to match the observed galaxy two-point
correlation function at $z=0$ in this cosmology. The simulation has a
box of length 141.3 $h^{-1}$ Mpc and contains $256^3$ dark matter
particles, each of mass of $1.4 \times 10^{10} h^{-1} M_{\odot}$.  We
identify halos in this simulation using the friends-of-friends (FOF)
algorithm with the standard linking length of 0.2, and then populate
them with galaxies according to the semi-analytic model.  We consider
only groups consisting of 10 particles or more, and so resolve dark
halos of mass $1.4 \times 10^{11} h^{-1} M_{\odot}$ or greater.
Sources in halos which are unresolved in the simulations can produce a
significant fraction of the total ionizing luminosity, according to
the semi-analytic models. To circumvent this problem, we add sources
in unresolved halos into the simulation in one of two ways. The first
method is to place the sources on randomly chosen dark matter
particles which do not belong to any resolved halo. An alternative
method is to place these sources completely at random within the
simulation volume. This makes the unresolved sources completely
unclustered and so is an interesting extreme case. As we will be
forced to construct toy models to determine which regions of the
simulation are ionized, the exact treatment of these unresolved halos
will not be of great importance. The number of unresolved halos added
to the simulation volume is determined from the Press-Schechter mass
function, multiplied by a correction factor of 0.7 to make it match
the low mass end of the N-body mass function in $\Lambda$CDM at $z=3$.

In order to calculate the correlations between ionized regions that
are needed to determine the temperature anisotropies induced in the
CMB, a simulation with at least the volume of this one is
required. Unfortunately, with present computing resources, this
excludes the possibility of an exact calculation of the shape and size
of the ionized regions, which would require much higher resolution,
and also the inclusion of gas dynamics and radiative transfer.
Therefore we have used five toy models to determine which regions of
the simulation are ionized, for a given distribution of ionizing
sources. These models cover a range of possibilities which is likely
to bracket the true case, and provide an estimate of the present
theoretical uncertainties.

For each model, we divide the simulation volume into $256^3$ cubic
cells, resulting in a cell size of $0.55 h^{-1}$ Mpc. As the gas
distribution is not homogeneous, the volume of gas ionized will depend
on the density of gas in the ionized region. We assume that the
ionizing luminosity from the galaxies in each halo all originates from
the halo centre, and that the total {\em mass} $M$ of gas ionized by
each halo is the same as it would be for an IGM which is uniform on
large scales, but with small-scale clumping $f_{\rm clump}$, as given
by eqn.~(\ref{ion_cgs}). We add to this the mass of any collisionally
ionized gas in the halo. We then calculate the {\em volume} of the
ionized region around each halo using $M={\overline n} _{\rm H} m_{\rm
H} V$, where ${\overline n} _{\rm H}$ is the mean IGM density within
the volume V.  We use several different toy models to calculate the
spatial distribution of ionized gas in the simulations. In all cases,
the total mass of hydrogen ionized is assumed to be the same as for a
homogeneous distribution with the specified clumping factor.

{\bf Model A (Growing front model)} Ionize a spherical volume around
each halo with a radius equal to the ionization front radius for that
halo assuming a large-scale uniform distribution of \HI. Since the
\HI\ in the simulation is \emph{not} uniformly distributed, and also
because some spheres will overlap, the ionized volume will not contain
the correct total mass of \HI. We therefore scale the radius of each
sphere by a constant factor, $f$, and repeat the procedure. This
process is repeated, with a new value of $f$ each time, until the
correct total mass of \HI\ has been ionized.

{\bf Model B (High density model)} In this model we ignore the
positions of halos in the simulation. Instead we simply rank the cells
in the simulation volume by their density. We then completely ionize
the gas in the densest cell. If this has not ionized enough \HI\ we
ionize the second densest cell. This process is repeated until the
correct total mass of \HI\ has been ionized.

{\bf Model C (Low density model)} As model B, but we begin by ionizing
the least dense cell, and work our way up to cells of greater and
greater density. This model mimics that of \scite{miralda00}.

{\bf Model D (Random spheres model)} As Model A but the spheres are
placed in the simulation entirely at random rather than on the dark
matter halos. By comparing to Model A this model allows us to estimate
the importance of the spatial clustering of dark matter halos.

{\bf Model E (Boundary model)} Ionize a spherical region around each
halo with a radius equal to the ionization front radius for that
halo. This may ionize too much or not enough \HI\ depending on the
density of gas around each source. We therefore begin adding or
removing cells at random from the boundaries of the already ionized
regions until the required mass of \HI\ is ionized.

\begin{figure*}
\begin{tabular}{cc}
Fully ionized box & Model A \\
\psfig{file=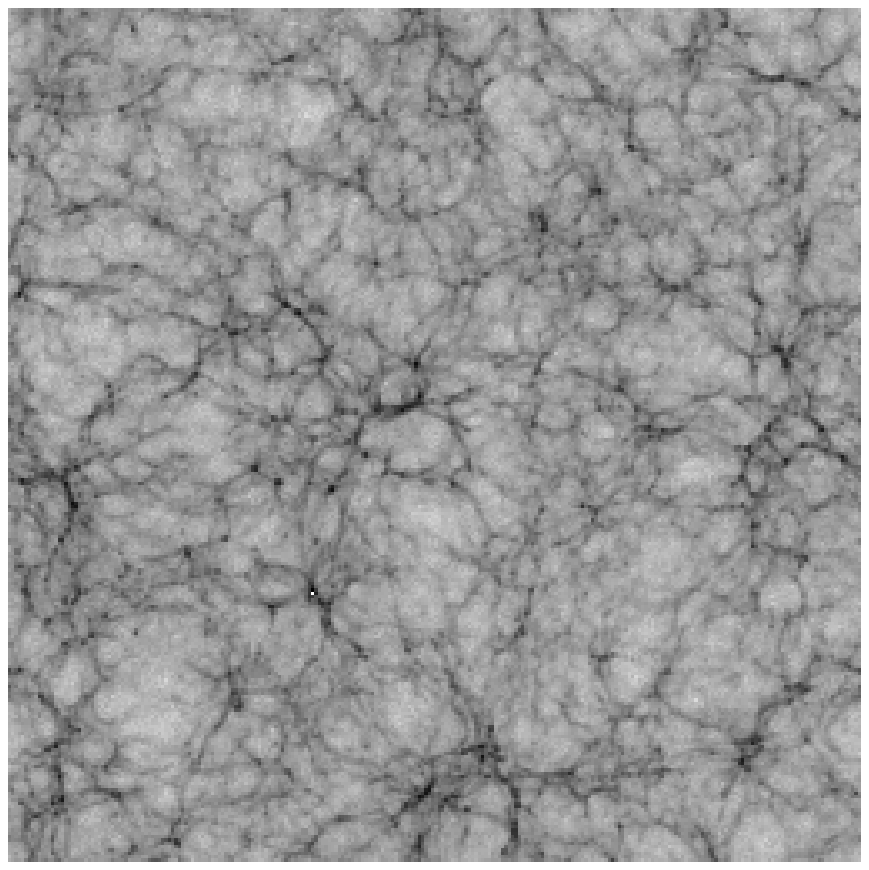,width=65mm} & \psfig{file=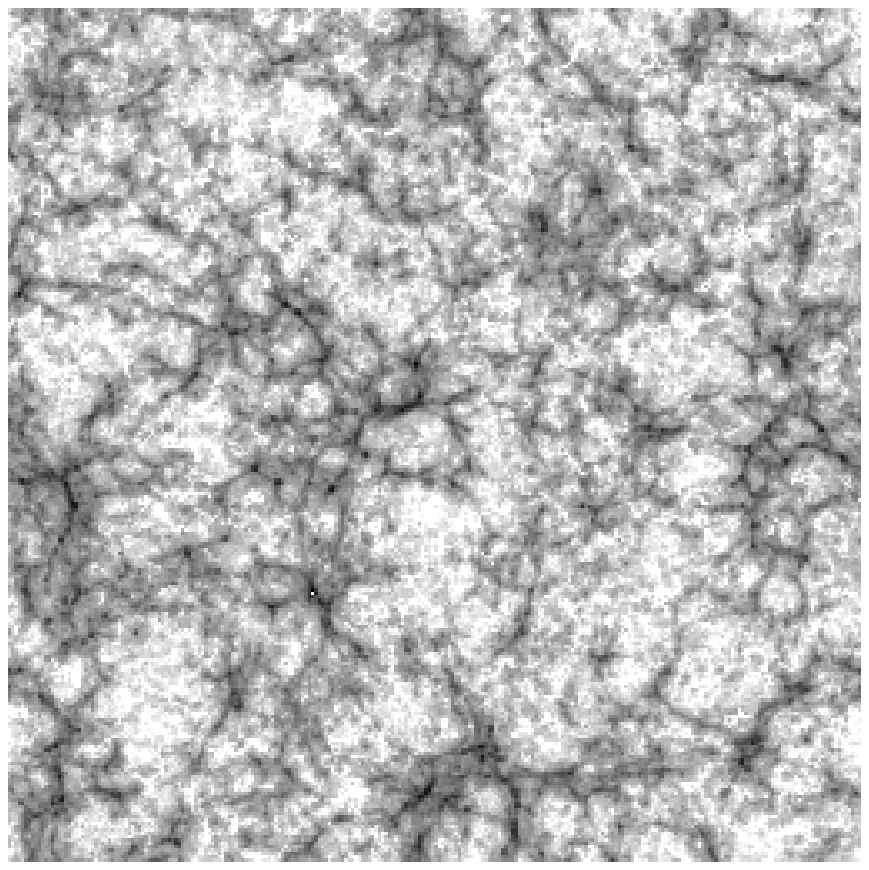,width=65mm} \\
Model B & Model C \\
\psfig{file=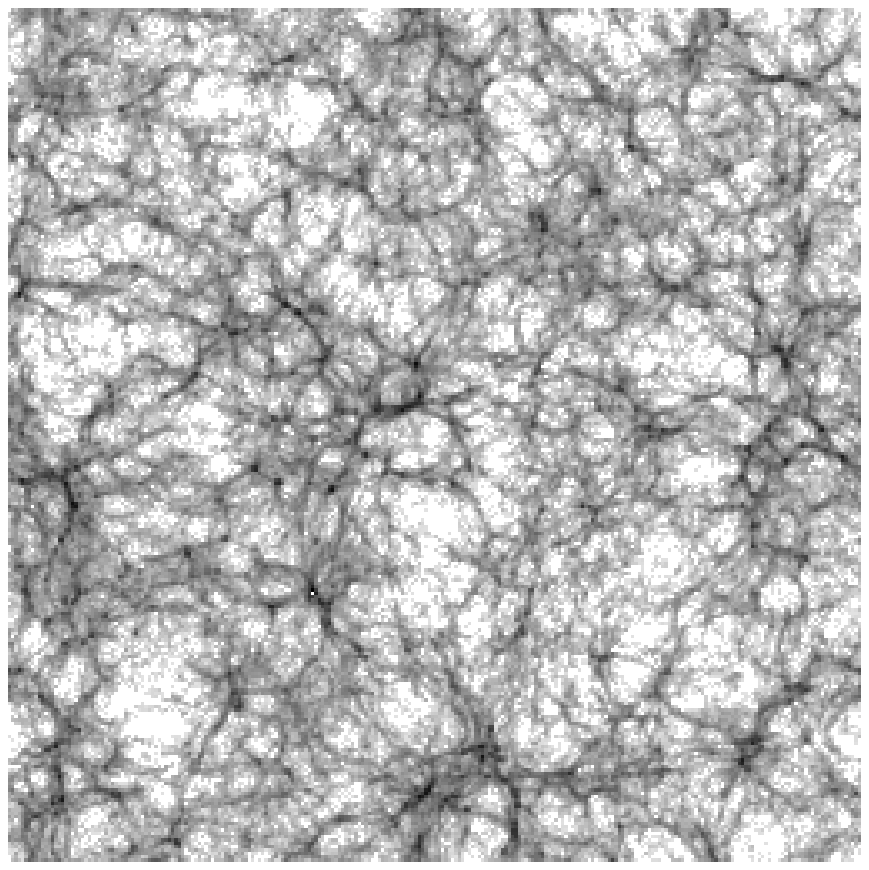,width=65mm} & \psfig{file=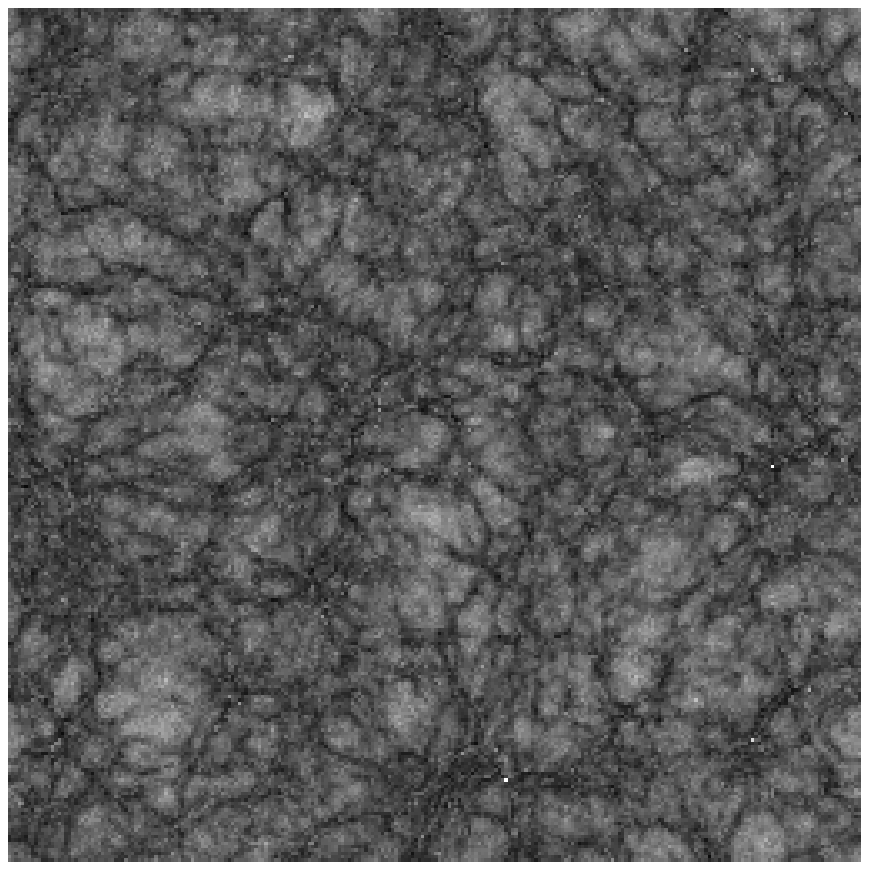,width=65mm} \\
Model D & Model E \\
\psfig{file=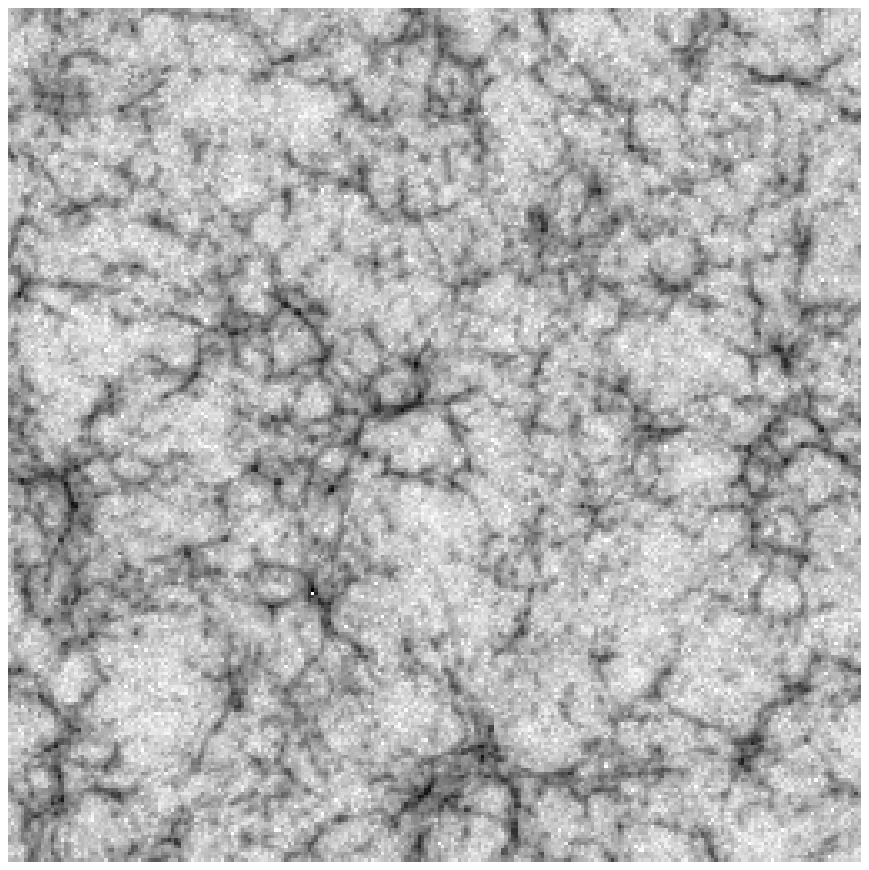,width=65mm} & \psfig{file=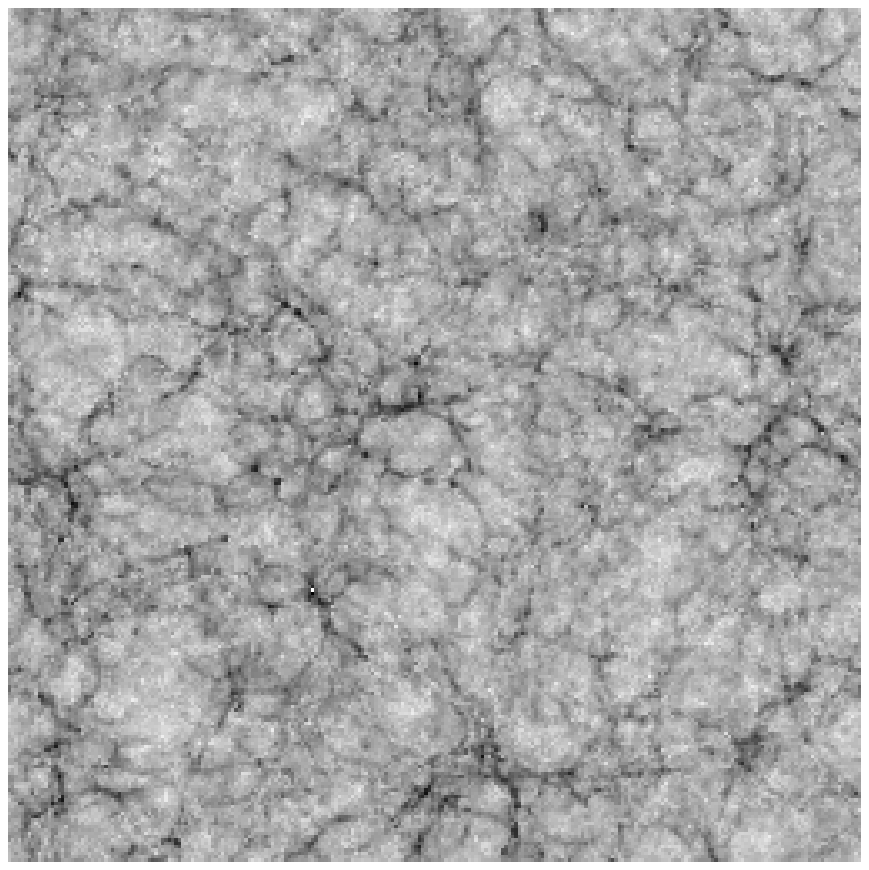,width=65mm} \\
\end{tabular}
\caption{The projected density of ionized gas in a slice
through the $\Lambda$CDM N-body simulation at $z=3$ shown as a
 greyscale image, with the densest regions being black.  The slice
 shown has dimensions of $141.3 \times 141.3 \times 8.0 h^{-1}$ Mpc.
 The total (i.e. ionized plus neutral) projected gas density is shown
 in the upper left hand panel.  The remaining panels show the
 projected density of ionized gas in Models A-E.}
\label{fig:ionregs}
\end{figure*}

Fig. \ref{fig:ionregs} shows six slices through the N-body
simulation. The top left slice shows the density of all gas (which is
assumed to trace the dark matter), whilst the other slices show only
the density of ionized gas. Model A shows particularly well the
correlated nature of the ionizing sources (due to the fact that
galaxies form in the high density regions of the dark matter), as the
densest regions of the simulations are the ones which have become most
highly ionized.

\begin{figure}
\psfig{file=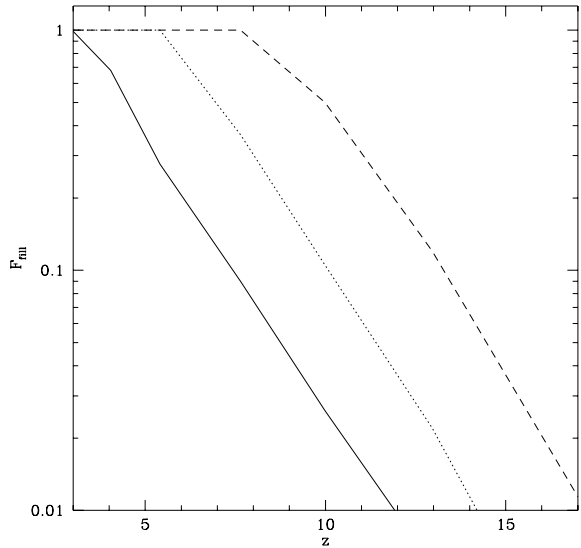,width=80mm}
\caption{Filling factors as a function of redshift in the N-body simulation. Lines are plotted for the `fixed' model with three different values of $f_{\rm esc,gas}$: 0.05 (solid line), 0.20 (dotted line) and 1.00 (short-dashed line).}
\label{fig:ffillsim}
\end{figure}

In Fig. \ref{fig:ffillsim} we show the filling factor in the N-body
simulation for the fixed $f_{\rm esc,gas}$ model with different values
of $f_{\rm esc,gas}$, for the case $f_{\rm clump}=f_{\rm clump}^{\rm
(halos)}$. The filling factors calculated from the simulation are
always less (for a given value of $f_{\rm esc,gas}$) than those
calculated in \S\ref{sec:ffresults} (see Fig. \ref{fig:Ffill}). This
is because the simulation contains fewer low mass dark matter halos
than predicted by the Press-Schechter theory, hence it contains fewer
ionizing sources.

\subsection{CMB Fluctuations}

The reionization of the IGM imprints secondary anisotropies on the CMB
through Thomson scattering off free electrons (see, for example,
\pcite{vishniac87,knox98,hu99}). These anisotropies result from the
spatially varying ionized fraction and from density and velocity
variations in the ionized IGM. The calculation of these secondary
effects involves correlation functions of density fluctuations and
velocity fields which are easily determined in our models. To predict
the form of these fluctuations, we first calculate the two-point
correlations between ionized gas over the redshift range 3 to 18,
assuming that gas in the IGM traces the dark matter density and
velocity. To do this, we use the $256^3$ grid of ionization fractions,
$x_{\rm e}$, described in
\S\ref{sec:spatdist}. We determine in each grid cell the value of
\begin{equation}
\zeta = \left[ {x_{\rm e} \left( 1+\delta\right) \over \langle
x_{\rm e} (1+\delta)\rangle} - 1 \right] v_{\rm los},
\end{equation}
where $\delta$ is the dark matter overdensity in the cell, $v_{\rm
los}$ is the component of the mean dark matter velocity in the cell
along the line of sight to a distant observer, and the averaging of
$x_{\rm e}(1+\delta )$ is over all cells in the simulation volume. The
dark matter density and velocity in each cell are estimated by
assigning the mass and velocity of each dark matter particle to the
grid using a cloud-in-cell algorithm.

We then compute the correlation function
\begin{equation}
\xi_{\zeta \zeta}(r) = \langle \zeta({\bf x}) \zeta({\bf x} + {\bf
r}) \rangle _{\bf x} .
\end{equation}
This correlation function is all that is needed to determine the
spectrum of fluctuations imprinted in the CMB by the reionization
process. A detailed description of how this spectrum is computed is
given in Appendix \ref{app:CMB}.

\begin{figure*}
\begin{tabular}{cc}
\psfig{file=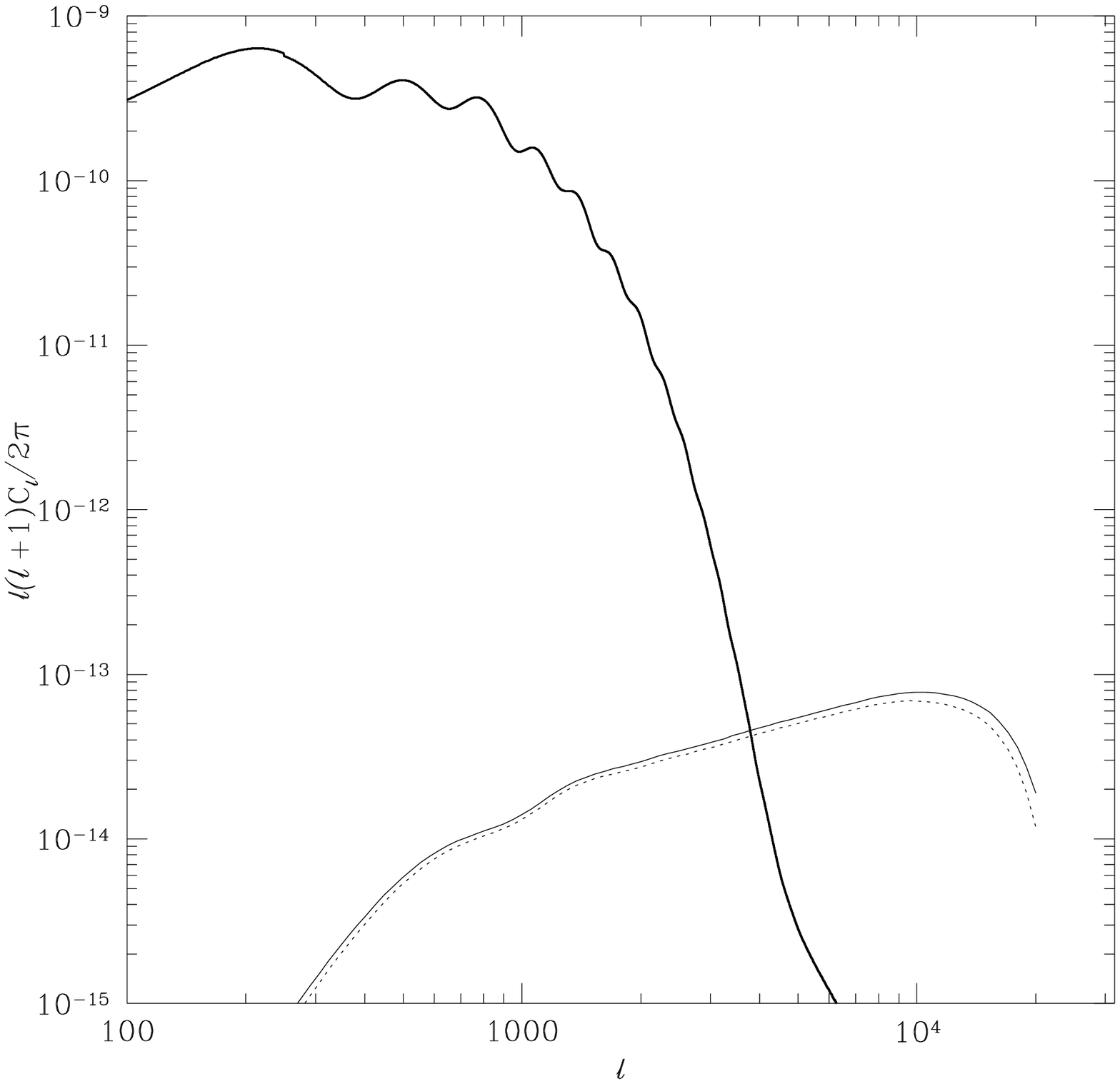,width=80mm} & \psfig{file=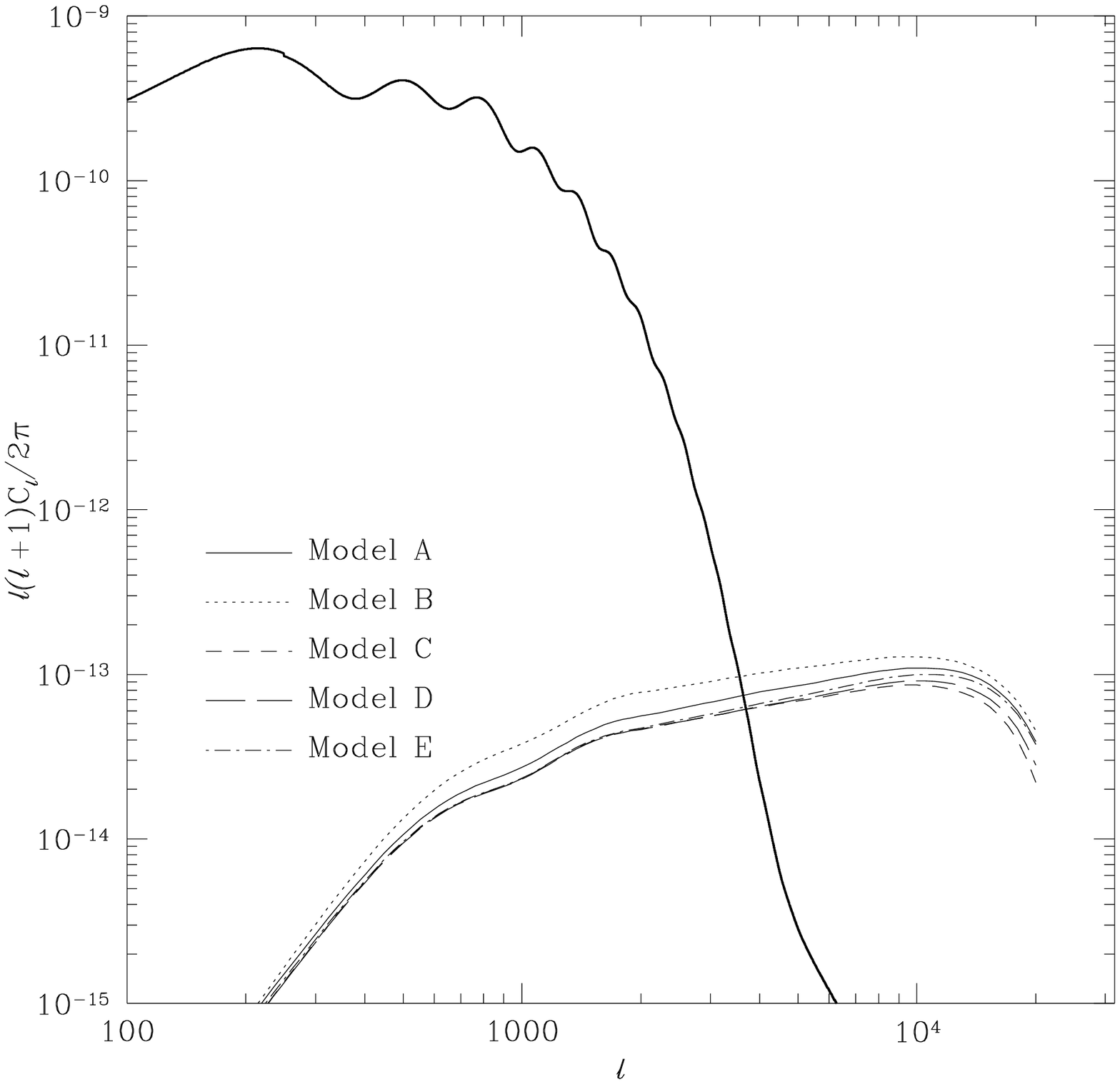,width=80mm}
\end{tabular}
\caption{The secondary CMB anisotropies generated measured from the
simulation. The left hand panel shows the results for Model E, with a
fixed gas escape fraction of 0.10. The solid line indicates a model in
which unresolved halos are placed on ungrouped particles, whilst the
dotted line shows a model with unresolved halos placed at random
within the simulation volume. In the right-hand panel we show the
results for $f_{\rm esc,gas}=1.00$ and with unresolved halos placed on
ungrouped particles. The lines show the results from the five
different models as indicated in the figure. In each case, the heavy
solid line shows the primary anisotropies in this cosmology.}
\label{fig:CLtest}
\end{figure*}

\begin{figure}
\psfig{file=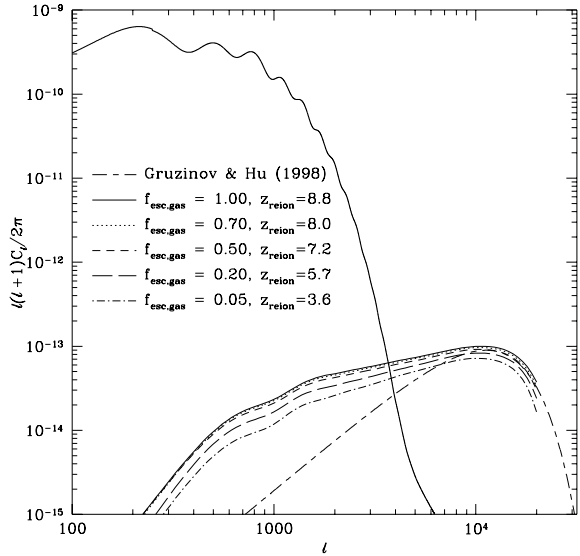,width=80mm}
\caption{The effect of varying the escape fraction, $f_{\rm esc,gas}$,
on the secondary CMB anisotropies. The curves shown are all computed
using Model E, with unresolved halos placed on ungrouped
particles. Gas escape fractions of 1.00, 0.70, 0.50, 0.20 and 0.05 are
shown as indicated in the figure. The redshift of reionization for
each model is also indicated. The heavy solid line shows the primary
CMB anisotropies.} 
\label{fig:CLfescvar}
\end{figure}

In Fig. \ref{fig:CLtest} we show the secondary CMB anisotropies
calculated as described above. The left-hand panel shows the results
for Model E with a fixed gas escape fraction of 0.10, for the cases
that unresolved halos are placed either on ungrouped particles (solid
line) or at random in the simulation volume (dotted line). The
clumping factor is $f_{\rm clump}=f_{\rm clump}^{\rm (halos)}$, as
will be used for all models considered in this section. The particular
choice of $f_{\rm clump}$ is not important for our conclusions about
the CMB fluctuations, since we will consider different values of
$f_{\rm esc,gas}$. The choice of placement scheme is seen to make
little difference to the results, the two curves differing by $\lsim
10\%$ for $250< \ell <5000$, with the difference growing to 40\% by
$\ell =20000$. The size of the grid cell used in our calculation of
the ionized gas correlation function corresponds to $\ell \approx
25000$ at $z=3$, and the turnover around $\ell = 10^4$ is simply due
to the cell size. As such, our method is unable to determine the form
of the CMB fluctuations at higher $\ell$.

The right-hand panel of Fig. \ref{fig:CLtest} shows the variations in
our estimates of $C_\ell$ which arise from using the five Models
A-E. Here the differences between the curves are larger, with Models B
and C differing by a factor of $\approx 2.5$ at $\ell = 10^4$. The
amplitude of the curves is affected by the strength of the
correlations present in each model (e.g. the ``high density'' model is
the most strongly correlated and has the highest amplitude, whilst the
``low density'' model has the weakest correlations and hence the
lowest amplitude). However, the shapes of the curves are all very
similar.

Figure \ref{fig:CLfescvar} examines the effect on the secondary
anisotropies of varying the escape fraction $f_{\rm esc,gas}$. The
trend is for increasing amplitude of anisotropy with increasing escape
fraction (which results in a higher reionization redshift). If,
however, we boost the number of photons produced by increasing $f_{\rm
esc,gas}$ above 1 (this of course being unphysical, but a simple way
of examining the effects of producing more ionizing photons), little
further increase in amplitude is seen.

The form expected for the CMB anisotropies produced by patchy
reionization has been calculated for a simple model by
\scite{gruzinov98}. In this model, reionization of the universe is
assumed to begin at some redshift, $z_{\rm i}$, and is completed
(i.e. the filling factor reaches unity) after a redshift interval
$\delta z$. Sources are assumed to appear at random positions in space
and to each ionize a spherical region of comoving radius $R$. Once
such an ionized region has appeared it remains forever. In this model,
the power $\ell^2C_\ell/2\pi$ is predicted to have the form of white
noise at small $\ell$, since the ionized regions are uncorrelated.

We compare the simple model of \scite{gruzinov98} with our own results
in Fig. \ref{fig:CLfescvar}.  Since in our model reionization has no
well-defined starting redshift, and ionized regions span a range of
sizes, we simply choose values of $R$, $z_{\rm i}$ and $\delta z$ in
order to match the two models at the peak in the spectrum (even though
the position of this peak in our results is an artifact of our
simulation resolution, we simply wish to demonstrate here the
difference in small $\ell$ slopes between our model and that of
\pcite{gruzinov98}). The chosen values of $R=0.85 h^{-1}$Mpc, $z_{\rm
i}=11$ and $\delta z=5$ are all plausible for the ionization history
and sizes of ionized regions seen in our model (the mean comoving size
of regions ranging from $1.4h^{-1}$Mpc at $z=3$ to $0.2h^{-1}$Mpc at
$z=18$). Note that \scite{knox98} calculate a somewhat different form
for the anisotropy spectrum for this same model, in which the
amplitude, $A$, is roughly half that found by
\scite{gruzinov98}, and the peak in the spectrum occurs at slightly
higher $\ell$. Despite these differences both \scite{gruzinov98}
and \scite{knox98} agree upon the general form of the spectrum
(sharp peak plus white noise at small $\ell$), and this is all we are
interested in here.

The $C_{\ell}$ declines much more rapidly as $\ell \rightarrow 0$ in
the \scite{gruzinov98} model than in ours.  Note that Model D,
the random sphere model, also shows the same behaviour as our other
models, indicating that it is not the correlated positions of the
ionizing sources in our model which produce the excess power at small
$\ell$. If we force all halos in our model to have equal ionized
volumes surrounding them, whilst retaining the same total filling
factor, we find that the excess power above the white noise spectrum
at small $\ell$ remains, so neither is the excess due to the range of
ionizing front radii, $R$, present in our model. This excess power can
therefore be seen to be due to the correlations in gas density and
velocity induced by gravity. In fact, if we repeat our calculations
but ignore correlations in the gas density field (i.e. we set $\delta
= 0$ everywhere) we find a CMB spectrum which has a slope for small
$\ell$ which is much closer to the \scite{gruzinov98} white-noise
slope, and which has an amplitude over five times lower than when
density correlations are included. The remaining differences between
our model and that of \scite{gruzinov98} in this case are due to
the correlated nature of ionized regions in our model.

\begin{figure}
\psfig{file=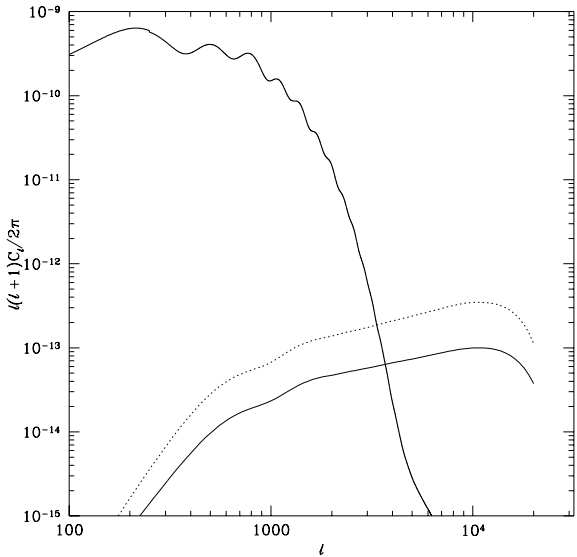,width=80mm}
\caption{The effect of varying $\Omega_{\rm b}$ on the secondary CMB anisotropies. The curves shown are all computed using Model E, with unresolved halos placed on ungrouped particles and a gas escape fraction of 1.0. The solid line shows $\Omega _{\rm b}=0.02$ whilst the dotted line shows $\Omega _{\rm b}=0.04$ (Model 1 in Table \protect\ref{tb:varmods}). The heavy solid line shows the primary CMB anisotropies.}
\label{fig:CLomegab}
\end{figure}

The amplitude of the secondary anisotropies also depends upon the
assumed value of $\Omega _{\rm b}$ (as this determines the optical
depth for electron scattering). The preferred value for our galaxy
formation model of $\Omega _{\rm b}=0.02$ is relatively low compared
to estimates based on light element abundances and big bang
nucleosynthesis. Fig. \ref{fig:CLomegab} shows the effect of
increasing $\Omega_{\rm b}$ to 0.04. If the evolution of the ionized
regions were the same in both models, we would expect the amplitude to
increase by a factor of four (since it is proportional to the square
of the baryon density). The evolution of ionized regions is actually
quite similar in the two cases, and so the factor of four increase is
seen. With this higher value of $\Omega_{\rm b}$, the secondary
anisotropies due to patchy reionization would be potentially
detectable above $\ell
\sim 3000$.

We note that a similar approach to computing the spectrum of secondary
anisotropies due to patchy reionization has been taken by
\scite{bruscoli00}. Using a different model of galaxy formation,
\scite{bruscoli00} grow spherical ionization fronts around dark
matter halos identified in an N-body simulation, and from these they
estimate the spectrum of CMB anisotropies produced. The simulations
employed by \scite{bruscoli00} have higher resolution (but much
smaller volume) than the GIF simulations used in our
work. \scite{bruscoli00} therefore do not have the problem of
locating unresolved halos in their simulation, but their calculation
of secondary anisotropies is restricted to smaller angular scales
(roughly $5\times 10^3 \lsim \ell \lsim 2\times 10^5$) compared to
ours.  Furthermore, \scite{bruscoli00} make some approximations in
calculating the anisotropies which we do not, ignoring variations in
the total IGM density, and assuming that the ionized fraction is
completely uncorrelated with the velocity field.
\scite{bruscoli00} consider only a single model for reionizing the
simulation volume. As we have shown, our five toy models for the
distribution of ionized regions lead to factors of 2--3 difference in
the secondary anisotropy amplitudes, indicating that the results are
not very sensitive to the model adopted for the distribution of
ionized regions or to the treatment of unresolved halos in the
simulation.
\scite{bruscoli00} carried out their calculations in
a different cosmology to ours, also with a different value of
$\Omega_{\rm b}$, but once these differences are taken into account,
their results seem reasonably consistent with ours.

\section{Discussion and Conclusions}
\label{sec:conc}

We have outlined an approach to studying the reionization of the
universe by the radiation from stars in high redshift galaxies. We
have focussed on the reionization of hydrogen, but the approach can be
generalised to study helium reionization (e.g. \pcite{giroux94}),
and also to include radiation from quasars. Our main conclusions are:

(i) Using a model of galaxy formation constrained by several
observations of the local galaxy population, enough ionizing photons
are produced to reionize the universe by $z=11.7$. This assumes that
all ionizing photons escape from the galaxies they originate in, and
that the density of the IGM is uniform. Reionization is delayed until
$z\approx 10.9$ in the case of a clumped IGM, in which gas falls into
halos with virial temperatures exceeding $10^4$K. Galaxies can
reionize such a clumped IGM by $z=5$ providing that, on average, at
least 4\% of ionizing photons can escape from the galaxies where they
are produced. In the case of a uniform IGM, an escape fraction of only
1.4\% is sufficient to reionize by $z=5$.

Using a physical model for the escape of ionizing radiation from
galaxies, in which photons escape through ``\HII\ chimneys'' ionized
in the gas layers in galaxy disks \cite[hereafter DS94]{doveshull94},
we predict reionization by $z=6.1$ for a uniform IGM or by $z=4.5$ for
a clumped IGM. Models which assume that all the gas in galaxy disks
remains neutral are unable to reionize even a uniform IGM by $z=0$.
Using alternative estimates of the IGM clumping factor from
\scite{gnedin97} or \scite{valageas99}, we find
reionization redshifts comparable with those found using our own
clumping model, i.e. in the range $z=$ 4.5--5.0 with the DS94 model
for the escape fraction.

(ii) Once the ionizing escape fraction and IGM clumping factor have
been specified, our estimates for the filling factor of ionized gas in
the IGM are reasonably robust, providing that we consider only models
which are successful in matching the H$\alpha$ luminosity function of
galaxies at $z=0$. By far the greatest remaining influences on the
ionized filling factor come from the value of the baryon fraction
$\Omega_{\rm b}$ and the prescription for feedback from
supernovae. However, we have shown that altering these parameters also
produces large changes in the $z=0$ H$\alpha$ luminosity function.

(iii) We combined our model for reionization with N-body simulations
of the dark matter distribution in order to predict the spectrum of
secondary anisotropies imprinted on the CMB by the process of
reionization. The {\it shape} of this spectrum is almost independent
of the assumptions about reionization, but the {\it amplitude} depends
on the spatial distribution of the ionized regions, the redshift at
which reionization occurs and the baryon fraction. We find
considerably more power in the anisotropy spectrum at small $\ell$
than predicted by models which do not account for the large-scale
correlations in the gas density and velocity produced by gravity.
Despite the uncertainty in the spatial distribution of ionized
regions, we are able to determine the amplitude of this spectrum to
within a factor of three for a given $\Omega_{\rm b}$ (the amplitude
being proportional to $\Omega_{\rm b}^2$). The results found by
\scite{bruscoli00} using a similar technique are reasonably
consistent with ours, once differences in $\Omega_{\rm b}$ and other
cosmological parameters are allowed for.

Detection of these secondary anisotropies, which would constrain the
reionization history of the Universe, would require fractional
temperature fluctuations of $\sim 10^{-7}$ to be measured on angular
scales smaller than several arcminutes. Although the Planck and MAP
space missions are unlikely to have sufficient sensitivity to observe
such anisotropies, the Atacama Large Millimeter Array is expected to
be able to measure temperature fluctuations of the level predicted at
$\ell \sim 10^4$ in a ten hour integration.

Previous studies of reionization have either used an approach similar
to our own, i.e. employing some type of analytical or semi-analytical
model (e.g. \pcite{haiman96,valageas99,chiu99,ciardi99}), or else have
used direct hydrodynamical simulations (e.g. \pcite{gnedin97}). While
the latter technique can in principle follow the detailed processes of
galaxy formation, gas dynamics and radiative transfer, in practice the
resolutions attainable at present do not allow such simulations to
resolve the small scales relevant to this problem. Furthermore, the
implementation of star formation and feedback in such models is far
from straightforward.

There are two main uncertainties in our approach, as in most others:
the fraction \fesctxt\ of ionizing photons that escape from galaxies,
and the clumping factor $f_{\rm clump}$ of gas in the IGM.  Future
progress depends on improving estimates of the effects of clumping
using larger gas dynamical simulations, on better modelling of the
escape of ionizing photons from galaxies, and on better understanding
of star formation and supernova feedback in high redshift objects.

\section*{Acknowledgements}
AJB and CGL acknowledge receipt of a PPARC Studentship and Visiting
Fellowship respectively. AN is supported by a grant from the Israeli
Science Foundation. NS is supported by the Sumitomo Foundation and
acknowledges the Max Planck Institute for Astrophysics for their warm
hospitality. This work was supported in part by a PPARC rolling grant,
by a computer equipment grant from Durham University and by the
European Community's TMR Network for Galaxy Formation and
Evolution. We acknowledge the Virgo Consortium and GIF for making
available the GIF simulations for this study. We are grateful to
Martin Haehnelt and Tom Abel for stimulating conversations. We also
thank Shaun Cole, Carlton Baugh and Carlos Frenk for allowing us to
use their galaxy formation model, and for advice on implementing
modifications in that model.

\appendix
\onecolumn

\section{Calculation of the escaping fraction}
\label{append:fesc}

\subsection{Escaping fraction in the DS94 model: Stars in mid-plane}

In the model of \scite[hereafter DS94]{doveshull94}, ionizing
photons escape from galactic disks through ``\HII\ chimneys'', which
are holes in the neutral gas layer ionized by OB associations. The OB
associations are assumed to lie in the disk mid-plane, and to have a
distribution of ionizing luminosities ${\rm d}N/{\rm d}S \propto
S^{-2}$ for $S_1<S<S_2$, $({\rm d}N/{\rm d}S){\rm d}S$ being the
number of associations with luminosities in the range $S$ to $S+{\rm
d}S$ \cite{kennedgar}. The gas is assumed to have a Gaussian vertical
distribution with scaleheight $h_{\mathrm z}$. The fraction of Lyc
photons escaping through chimneys on both sides of the disk at radius
$r$ is
\cite[eqn. 24]{doveshull94}
\renewcommand{\arraystretch}{2}
\begin{equation}
f_{\rm esc,gas} = \left\{ \begin{array}{ll} 0 & \hbox{if } S_{\rm m}
\geq S_2 \\ \left. \left[ \ln \left( {S_2 \over S_{\mathrm m}} \right)
+ {9 \over 2} \left({S_{\rm m}\over S_2}\right)^{1/3} - {S_{\rm m}
\over 2 S_2} - 4
\right] \right/ \ln \left( {S_2 \over S_1} \right) & \hbox{if } S_1 \leq S_{\rm m} < S_2 \\ 1 + \left. \left[ {9 \over 2} \left\{ \left({S_{\rm m}\over S_2}\right)^{1/3} -  \left({S_{\rm m}\over S_1}\right)^{1/3}\right\} - {1 \over 2} \left\{ {S_{\rm m} \over S_2}- {S_{\rm m} \over S_1} \right\}
\right] \right/ \ln \left( {S_2 \over S_1} \right) & \hbox{if } S_{\rm m} < S_1, \end{array} \right.
\label{eq:ds}
\end{equation}
where  $S_{\mathrm m}$ is defined as
\begin{equation}
S_{\rm m}(r) = \pi ^{3/2} n_0^2 \exp \left( -2r/r_{\rm disk}\right) h_{\mathrm z}^3 \alpha _{\mathrm H}^{(2)},
\end{equation}
Here $\alpha _{\mathrm H}^{(2)}$ is the recombination coefficient for
hydrogen for recombinations to all energy levels except the
first, and we have assumed an exponential disk with radial scalelength
$r_{\rm disk}$, so that the hydrogen gas density is
\begin{equation}
n(r,z)=n_0\exp(-r/r_{\rm disk}-z^2/2h_{\rm z}^2),
\label{eq:diskdens}
\end{equation}
$r$ and $z$ being the usual cylindrical polar coordinates.

Since $S_{\rm m}$ varies throughout the galactic disk we average the
escape fraction over the entire disk, assuming that the local rate of
star formation is proportional to the column density of the disk
\cite{kennicutt89,kennicutt97} and that $h_{\rm z}$ is constant with
radius.

\begin{figure}
\hspace{45mm}\psfig{file=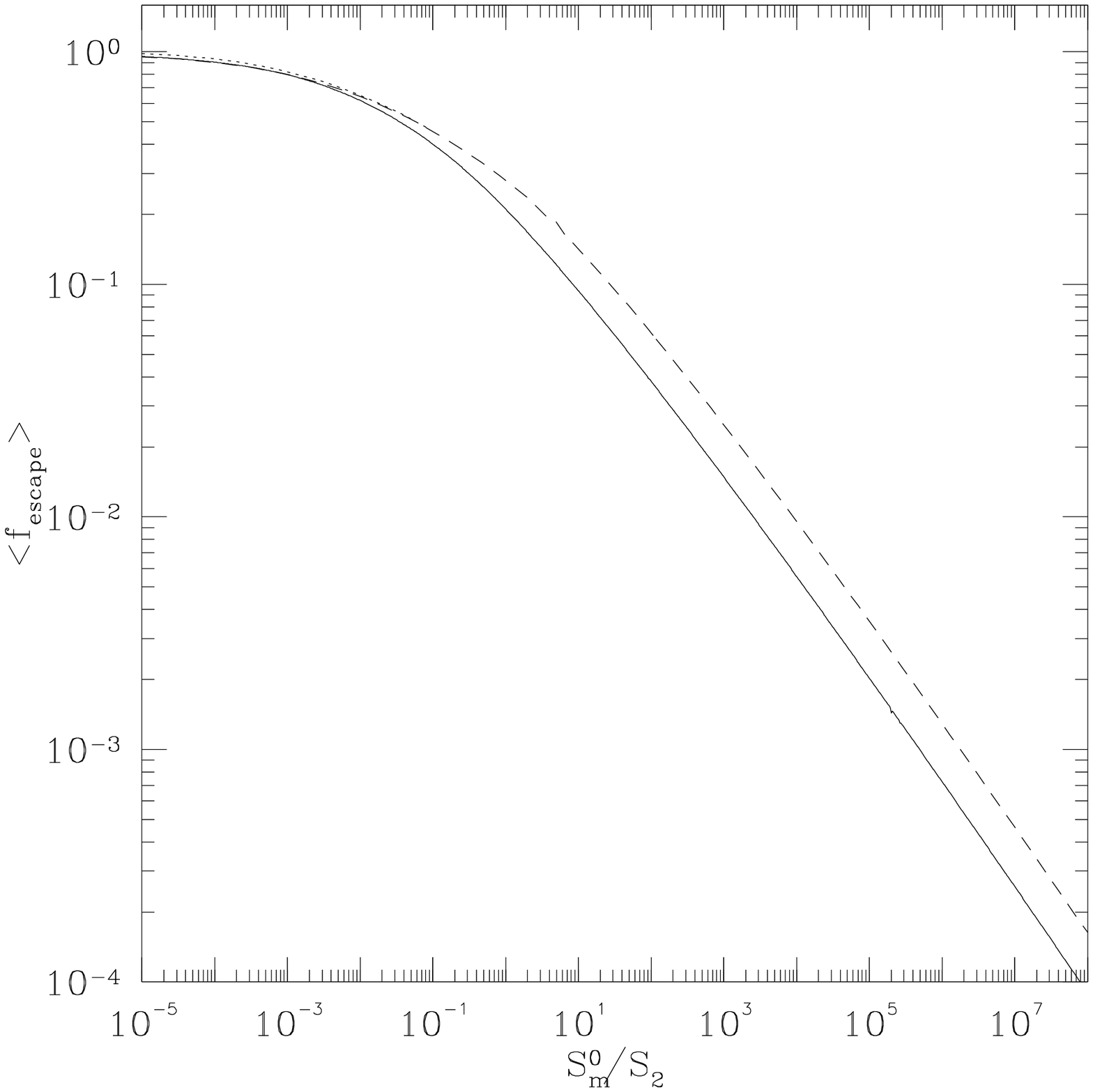,width=80mm} \caption{The
average escape fraction for a galactic disk in the DS94 model with
OB associations in the disk mid-plane (solid line) and distributed
as the cold gas (dashed line). Both models assume $S_2/S_1=1000$.
The dotted line (visible just above the solid line at small values
of $S_{\rm m}^0/S_2$) shows the effects of accounting for radial
variations in the gas density in the model where OB associations
are distributed as the gas. This line only differs noticeably from
the dashed line at the lowest values of $S^0_{\mathrm m}/S_2$.}
\label{fig:fescDS}
\end{figure}

The fraction of all ionizing photons produced by the galaxy which can
escape into the IGM is then given by,
\begin{equation}
f_{\rm esc,gas} = \left\{ \begin{array}{ll} \left. \left[ \left\{{352\over 225}  - {4\over 15} \ln\left({S_2\over S_{\rm m}^0}\right) \right\} \left({S_2 \over S_{\rm m}^0}\right)^{1/2} - \left\{{352\over 225}  - {4\over 15} \ln\left({S_1\over S_{\rm m}^0}\right) \right\} \left({S_1 \over S_{\rm m}^0}\right)^{1/2}\right] \right/ \ln\left({S_2\over S_1}\right) & \hbox{if } S_{\rm m}^0 \geq S_2 \\ \left. \left[ \left\{ {4\over 15} \ln \left( {S_1\over S_{\rm m}^0}\right) - {352\over 225} \right\} \left( {S_1\over S_{\rm m}^0} \right)^{1/2} - {1\over 18} {S_{\rm m}^0 \over S_2} + {81\over 50} \left({S_{\rm m}^0\over S_2}\right)^{1/3} + \ln\left({S_2\over S_{\rm m}^0}\right) \right] \right/ \ln \left({S_2\over S_1}\right) & \hbox{if } S_1 \leq S_{\rm m}^0 < S_2 \\ 1 + \left. \left[ {81\over 50} \left\{ \left({S_{\rm m}^0 \over S_2}\right)^{1/3} - \left({S_{\rm m}^0\over S_1}\right)^{1/3}\right\} - {1\over 18} \left\{ {S_{\rm m}^0 \over S_2} - {S_{\rm m}^0\over S_1} \right\} \right] \right/ \ln \left( {S_2 \over S_1}\right) & \hbox{if } S_{\rm m}^0 < S_1, \end{array} \right.
\label{eq:fescr}
\end{equation}
where $S^0_{\rm m}$ is the value of $S_{\rm m}$ calculated for
$r=0$. In Figure \ref{fig:fescDS} we show this average escape fraction
as a function of the ratio $S^0_{\mathrm m}/S_2$ for $S_2/S_1=1000$.

Our model of galaxy formation calculates the radial scale length of
each galaxy's disk and also the mass of cold gas present in that disk,
and we assume that $h_{\mathrm z}/r_{\rm disk}$ is constant. We can
therefore determine $n_0$ and the ratios $S^0_{\rm m}/S_2$ and
$S^0_{\rm m}/S_1$. Hence $f_{\rm esc,gas}$ can be found using
eqn.~(\ref{eq:fescr}).

\subsection{Escaping fraction in DS94 model: Stars tracing gas}

In the DS94 model OB associations are assumed to lie in the midplane
of the galaxy disk. If instead OB associations are spread throughout
the gas layer, having the same vertical distribution as the cold gas,
then the resulting escape fraction will be higher than that in the
DS94 model. We assume the same density profile as before, given by
eqn.(\ref{eq:diskdens}).  Consider an OB association emitting $S$
ionizing photons per second, at position $(r,z)$ in the disk. We make
the assumption (as did DS94) that the radial variations in density can
be ignored for calculating the escape fraction at radius $r$ (which
will be a valid assumption provided the size of the
\HII\ region formed is much less than $r_{\rm disk}$). In order for
any photons emitted into a cone of solid angle ${\rm d}\Omega$ which
makes an angle $\theta$ with the $z$-axis to escape the galaxy, the
emission rate of photons into this cone must exceed the total
recombination rate in the cone. This occurs for an ionizing luminosity
$S_{\rm req}(\theta)$, where,
\begin{equation}
S_{\rm req}(\theta) {{\rm d}\Omega \over 4\pi} = n_0^2 \alpha_{\rm H}^{(2)} \exp \left( -2{r \over r_{\rm disk}} \right) \int_0^\infty l^2 \exp \left( - {(z+l\cos\theta)^2 \over h_{\rm z}^2} \right) {\rm d}l {\rm d}\Omega ,
\end{equation}
which can be written as $S_{\rm req}^\pm(\theta) = \pm S_{\rm
req}^{0,\pm}/ \cos^3\theta$ ($S_{\rm req}^+$ is the solution for
$\cos\theta > 0$ and $S_{\rm req}^-$ is the solution for $\cos\theta <
0$), where
\begin{equation}
S_{\rm req}^{0,\pm} = S_{\rm m}^0 {\rm e}^{-2r/r_{\rm disk}} \left\{ \left[ 1 \mp {\rm erf} \left( {z \over h_{\rm z}} \right) \right] \left( 1 + {2 z^2 \over h_{\rm z}^2}\right) - {2 \over \sqrt{\pi}}{z \over h_{\rm z}} \exp \left( -{z^2 \over h_{\rm z}^2} \right) \right\}.
\end{equation}
This defines two critical angles, $\cos \theta_{\rm c}^\pm(S)=\pm (S_{\rm
req}^{0,\pm}/S)^{1/3}$, such that photons can escape the galaxy only if $\theta
< \theta_{\rm c}^+(S)$ or $\theta > \theta_{\rm c}^-(S)$.

The total escaping fraction from this OB association is then given by,
\begin{eqnarray}
f_{\rm esc,gas}(S) & = & {1 \over 2 S} \left[ \int_{\cos
\theta_{\rm c}^+}^1 \left[ S - S_{\rm req}^+(\theta)\right] \: {\rm d}(\cos
\theta) + \int^{\cos \theta_{\rm c}^-}_{-1} \left[ S - S_{\rm req}^-(\theta) \right]
\: {\rm d}(\cos \theta) \right] \\ f_{\rm esc,gas}(S) & = & 1 - {3
\over 4}\left({S_{\rm req}^{0,+} \over S}\right)^{1/3} - {3 \over
4}\left({S_{\rm req}^{0,-} \over S}\right)^{1/3}+{S_{\rm
req}^{0,+} \over 4 S}+{S_{\rm req}^{0,-} \over 4
S}.
\end{eqnarray}
Averaging this escape fraction over the assumed OB association
luminosity function then gives a mean escape fraction of
\begin{equation}
f_{\rm esc,gas} = \left\{ \begin{array}{ll} 0 & \hbox{if } S_2 > S_{\rm m}^0 {\rm e}^{-2r/r_{\rm disk}} \\ \begin{array}{r} \left. \left[{1\over 2}\ln\left({S_2^2\over S_{\rm req}^{0,+}S_{\rm req}^{0,-}}\right) + {9\over 4} \left\{ \left({S_{\rm req}^{0,+}\over S_2}\right)^{1/3}+\left({S_{\rm req}^{0,-}\over S_2}\right)^{1/3}-2\right\}\right. \right. \\ \left. \left. - {1\over 4}\left\{ {S_{\rm req}^{0,+}\over S_2}+{S_{\rm req}^{0,-}\over S_2}-2\right\}\right]\right/\ln\left({S_2\over S_1}\right) \end{array} & \hbox{if } S_1 < S_{\rm m}^0 {\rm e}^{-2r/r_{\rm disk}} \leq S_2 \\ \begin{array}{r} 1 + \left .\left[ {9\over 4} \left\{ \left({S_{\rm req}^{0,+}\over S_2}\right)^{\frac{1}{3}}-\left({S_{\rm req}^{0,+}\over S_1}\right)^{\frac{1}{3}}+\left({S_{\rm req}^{0,-}\over S_2}\right)^{\frac{1}{3}}-\left({S_{\rm req}^{0,-}\over S_1}\right)^{\frac{1}{3}} \right\}\right.\right. \\ \left.\left. - {1 \over 4}\left\{ {S_{\rm req}^{0,+}\over S_2}-{S_{\rm req}^{0,+}\over S_1}+{S_{\rm req}^{0,-}\over S_2}-{S_{\rm req}^{0,-}\over S_1}\right\} \right] \right/ \ln\left({S_2\over S_1}\right) \end{array} & \hbox{if } S_{\rm m}^0 {\rm e}^{-2r/r_{\rm disk}} \leq S_1 \end{array} \right.
\end{equation}
This expression is then averaged over the galaxy disk, assuming a star
formation rate proportional to the local gas density, to derive the
mean escaping fraction for the entire galaxy. This must be done
numerically

Although we have ignored radial variations in the density of the gas
when computing the escaping fraction from a single OB association, we
find by numerical solution that these variations make only a very
small difference to the value of $f_{\rm esc,gas}$, and then only for
small $S_{\rm m}^0/S_2$ (see Fig. \ref{fig:fescDS}).

\subsection{Escaping fraction in the DSGN98 model}

In the model of \scite[DSGN98]{devriendt99} the stars producing
the Lyman continuum photons are assumed to be uniformly mixed with the
gas in the galaxy, which is distributed in an exponential disk. All of
the hydrogen in the galaxy is assumed to be in the form of \HI,
allowing the optical depth for ionizing photons to be calculated.

\begin{figure}
\hspace{45mm}\psfig{file=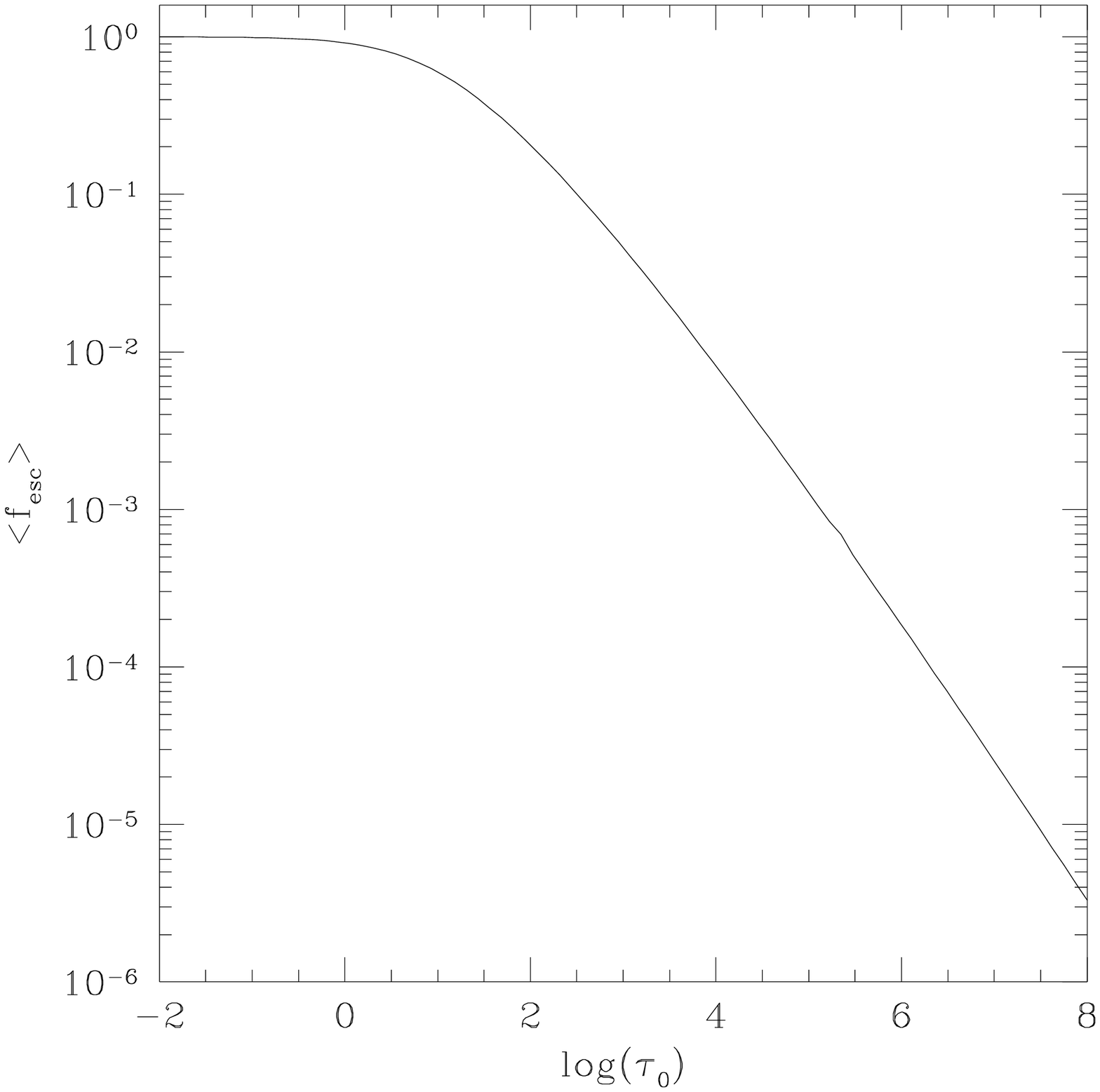,width=80mm}
\caption{Mean escape fraction for a galaxy disk in the DSGN98 model as a function of $\tau _0$.}
\label{fig:DSGNfesc}
\end{figure}

We have calculated the escaping fraction in this model exactly, using
the density profile given by eqn. (\ref{eq:diskdens}). As in the case
of the DS94 model with stars mixed uniformly with the gas, we begin by
finding the escaping fraction as a function of position $(r_0,z_0)$
and line of sight $(\theta,\phi)$. For the DSGN98 model we therefore
find the total optical depth in neutral hydrogen along the line of
sight, which is
\begin{equation}
\tau_(r_0,z_0,\theta,\phi) = \sigma _{\rm H{\sc i}} \int^{\infty}_0 n(r,z) {\rm d}l,
\end{equation}
where $\sigma _{\rm H{\sc i}}=6.3 \times 10^{-18} {\rm cm}^{-2}$ is
the cross-section for hydrogen ionization at 912\AA. The quantity
$\exp(-\tau)$ is then averaged over all $r_0$, $z_0$, $\theta$ and
$\phi$, assuming that the star formation rate is proportional to the
local gas density, to obtain the final escaping fraction. The average
escape fraction for the entire galaxy depends only on the quantity
$\tau_0=n_0 r_{\rm disk} \sigma_{\rm H}$ for a given value of $h_{\rm
z}/r_{\rm disk}$, and is shown in Fig. \ref{fig:DSGNfesc} for $h_{\rm
z}/r_{\rm disk}=0.1$.

\subsection{Escaping fraction in starbursts}

In the case of a burst of star formation triggered by a major merger,
we use the same $f_{\rm esc,gas}$ as for quiescent star formation in
the the case where $f_{\rm esc,gas}$ is assumed fixed, but in the DS94
and DSGN98 models we estimate the escape fraction by assuming the
burst has an approximately spherical geometry, throughout which star
formation proceeds uniformly. We assume a sphere of uniform hydrogen
number density, $n$, given by
\begin{equation}
n = {3 M_{\rm gas} \over 4 \pi r_{\rm burst}^3 1.4 {\rm m}_{\rm H}},
\end{equation}
where $M_{\rm gas}$ is the mass of cold gas in the burst, $r_{\rm
burst}$ is the radius of the region in which the burst
occurs, and ${\rm m}_{\rm H}$ is the mass of a hydrogen atom. The
factor of 1.4 accounts for the presence of helium in the gas. We
assume also that photons escape only from an outer shell of
thickness $l$, within which the optical depth is less than 1.
Therefore,
\begin{equation}
n l \sigma_{\rm H{\sc i}} \approx 1,
\end{equation}
where $\sigma_{\rm H{\sc i}}$ is the cross section for hydrogen
ionization. The escape fraction is simply the fraction of the sphere's
volume in this shell, i.e.
\begin{equation}
f_{\rm esc,gas} \approx {4 \pi r_{\rm burst}^2 l \over 4 \pi /3 r_{\rm burst}^3}.
\end{equation}
Substituting for $l$ then gives
\begin{eqnarray}
f_{\rm esc,gas} & = & {3 \over r_{\rm burst} n \sigma_{\rm H{\sc i}}} \nonumber \\
 & = & {4 \pi r_{\rm burst}^2 1.4 {\rm m}_{\rm H} \over M_{\rm gas} \sigma_{\rm H{\sc i}}}.
\end{eqnarray}
We take $r_{\rm burst}$ to be equal to $0.1 r_{\rm bulge}$ where
$r_{\rm bulge}$ is the half-mass radius of the bulge formed by the
merger. This choice is motivated by observational fact which shows
that starburst activity is usually confined to the nuclear region, the
size of which is much smaller than that of the galaxy as a whole
(e.g. \pcite{sanders96} and references therein). \scite{ricotti99}
have carried out more elaborate calculations of escaping fractions
from spherical galaxies. However, their results are applicable to gas
in hydrostatic equilibrium with an NFW dark matter profile and so are
not well suited to the case of starbursts.

The star formation rate in the burst is assumed to decline
exponentially, with an e-folding time equal to $f_{\rm dyn}$ times
the bulge dynamical time. Unless noted otherwise, we assume $f_{\rm
dyn}=1$ in all models. As the burst proceeds the mass of cold gas
present, $M_{\rm gas}$, declines as it is turned into stars. The
escape fraction, $f_{\rm esc,gas}$, therefore increases during the
burst, reaching unity as the amount of gas present drops to zero.
However, as the star formation rate is declining exponentially
during the burst only a small fraction of photons are produced
whilst $f_{\rm esc,gas}$ is high.

\section{Calculation of Clumping Factor}
\label{append:fclump}
\label{sec:clumphalo}

To estimate the clumping factor of the photoionized IGM, we make the
simplifying assumption that gas in the universe can be split into
three components --- that which has fallen into dark matter halos and
is collisionally ionized or is part of a galaxy, that which has fallen
into dark matter halos and is \emph{not} collisionally ionized, and
that which has remained outside halos and is smoothly distributed. The
first component makes no contribution to the clumping factor. We
define the clumping factor as
\begin{equation}
f_{\rm clump}={\langle \rho_{\rm IGM}^2 \rangle \over \bar{\rho}_{\rm IGM}^2}=
{\langle \rho_{\rm IGM}^2 \rangle \over f_{\rm IGM}^2 \bar{\rho}^2},
\end{equation}
where $\rho_{\rm IGM}$ is the IGM gas density at any point in the
universe (i.e. it does \emph{not} include contributions from
collisionally ionized gas or galaxies), $\bar{\rho}_{\rm IGM}=f_{\rm
IGM} \bar{\rho}$ is the mean density of gas in the IGM and
$\bar{\rho}$ is the mean density of all gas in the universe (here
$f_{\rm IGM}$ is the fraction of the total mass of gas in the universe
which resides in the IGM, as defined in \S\ref{sec:IGM}).

Let $f_{\rm m,clumped}$ be the fraction of mass in
halos above the Jeans halo mass, $M_{\rm J}$, as calculated from the
Press-Schechter mass function for example. These halos occupy a
fraction of the volume of the universe given by $f_{\rm
v,clumped}=f_{\rm m,clumped}/\Delta_{\rm vir}$. Here, $\Delta_{\rm
vir}$ is the mean density within the virial radius of a halo in units
of the mean density of the Universe. The smooth component of gas is
assumed to uniformly fill the region outside halos with $M>M_{\rm J}$,
and so has density
\begin{equation}
\rho _{\rm smooth} =\bar{\rho} f_{\rm m,smooth}/f_{\rm v,smooth},
\end{equation}
where $f_{\rm m,smooth}=1-f_{\rm m,clumped}$ is the mass fraction of
gas in this smooth component, and $f_{\rm v,smooth}=1-f_{\rm
v,clumped}$ is the fraction of the volume of the universe that it occupies.

Consider next the non-collisionally ionized gas in a \emph{single} dark
matter halo. Averaging over the volume of this one halo we obtain
\begin{equation}
\langle \rho_{\rm clumped}^2 \rangle = f_{\rm int} (1-f_{\rm gal})^2
(1-x_{\rm H})^2 \Delta_{\rm vir}^2 \bar{\rho}^2, 
\end{equation}
where $f_{\rm gal}$ is the fraction of the baryons which have become
part of galaxies within the halo, $x_{\rm H}$ is the ionized fraction
for the hydrogen in the halo gas assuming collisional ionization
equilibrium (which we take from the calculations of
\pcite{sutherland93}), and $f_{\rm int}$ is a factor of order unity
which depends on the shape of the halo gas density profile and is
given by
\begin{equation}
f_{\rm int}={\int_0^{r_{\rm vir}} \rho^2(r) r^2 {\rm d}r \over
\int_0^{r_{\rm vir}} \bar{\rho}_{int}^2 r^2 {\rm d}r }.
\label{eq:fint}
\end{equation}
Here $r_{\rm vir}$ is the virial radius of the halo, $\rho(r)$ is the
density profile of the diffuse gas in the halo, and $\bar{\rho}_{int}$
is the mean density of this gas within the virial radius. We ignore
any dependence of the density profile of the gas in the halo on the
fraction which has cooled to form galaxies. Our results should be
insensitive to this assumption, as $f_{\rm gal}\ll 1$ in halos where
$x_{\rm H}$ is significantly less than unity.

To find the contribution of gas in halos to the clumping factor, we
integrate the above expression over all halos more massive than
$M_{\rm J}$, weighting by the volume for each halo. Adding the
contribution from the smooth component, we then obtain
\begin{equation}
f_{\rm clump} = {f_{\rm m,smooth}^2 \over f_{\rm v,smooth} f_{\rm IGM}^2} + {f_{\rm int} \Delta_{\rm vir} \over f_{\rm IGM}^2} \int_{M_{\rm J}}^\infty \langle (1-f_{\rm gal})^2 \rangle (1-x_{\rm H})^2 {M_{\rm halo} \over \rho_{\rm c} \Omega_0} {{\rm d}n \over {\rm d}M_{\rm halo}} {\rm d}M_{\rm halo},
\end{equation}
where we have used that fact that the comoving volume of a dark matter
halo of mass $M_{\rm halo}$ is $M_{\rm halo}/(\Delta_{\rm vir}
\Omega_0 \rho_{\rm c})$ ($\rho_{\rm c}$ being the critical density of
the universe at $z=0$). Here $\langle (1-f_{\rm gal})^2 \rangle$ is
averaged over all halos of mass $M_{\rm halo}$ in our model of galaxy
formation.

We determine $M_{\rm J}$ by finding the mass of a dark matter halo which
has a potential well deep enough that it can just hold onto reionized gas.
This gives us the minimum mass halo within which gas collects. For the
halo to just retain its gas,
\begin{equation}
{{\rm d}P \over {\rm d}r} = {{\rm G} M_{\rm J} \over r_{\rm vir}^2} \rho (r_{\rm vir}),
\end{equation}
where $r_{\rm vir}$ is the virial radius of the halo
 and $P$ is the gas pressure. We approximate this as
\begin{equation}
{P \over r_{\rm vir}} \approx {{\rm G} M_{\rm J} \over r_{\rm vir}^2} \rho (r_{\rm vir}),
\end{equation}
and using the ideal gas law this becomes
\begin{equation}
{{\rm k}_{\rm B} T \over \mu {\rm m}_{\rm H}} \approx {{\rm G} M_{\rm J} \over r_{\rm vir}} = {4 \pi \over 3} {\rm G} r^2_{\rm vir} \rho_{\rm c} \Omega_0 \Delta_{\rm vir} (1+z)^3,
\end{equation}
where we have used the relation $M_{\rm J}=4 \pi \rho_{\rm c}
\Omega_0 (1+z)^3 \Delta_{\rm vir} r_{\rm vir}^3/3$. The virial
radius is therefore
\begin{equation}
r_{\rm vir} = \left({3 \over 4 \pi} {{\rm k}_{\rm B} T \over
{\rm G} \mu {\rm m_{\rm H}} \rho_{\rm c} \Omega_0 \Delta_{\rm vir}}
\right)^{1/2} (1+z)^{-3/2},
\end{equation}
and the minimum halo mass in which gas is retained is
\begin{equation}
M_{\rm J} = {4 \pi \over 3} \rho_{\rm c} (1+z)^3 \Omega_0 \Delta_{\rm vir} r_{\rm vir}^3.
\end{equation}

We evaluate $f_{\rm int}$ for the case of an isothermal profile with
core radius $r_{\rm c}$:
\begin{equation}
\rho(r) \propto {1 \over r^2 + r_{\rm c}^2}.
\end{equation}
The simulations of galaxy clusters by \scite{navarro95} and
\scite{eke98} show that the gas density profile is well
described by this form. Substituting this in eqn.~(\ref{eq:fint}), we
find
\begin{equation}
f_{\rm int} = {1 \over 6} \left( { r_{\rm vir} \over r_{\rm c}} \right)^3  \left[ {r_{\rm vir} \over r_{\rm c}} - \arctan {r_{\rm vir} \over r_{\rm c}} \right]^{-2} \left[ \arctan {r_{\rm vir} \over r_{\rm c}} - {r_{\rm vir} \over r_{\rm c}} \left( 1 + {r_{\rm vir}^2 \over r_{\rm c}^2} \right)^{-1} \right].
\end{equation}
For a typical value of $r_{\rm vir}/r_{\rm c}=10$, we therefore find
$f_{\rm int}=3.14$.

\section{The Spectrum of CMB Secondary Anisotropies}
\label{app:CMB}

In this paper, we concentrate on the kinematic Sunyaev-Zel'dovich
effect which is induced by the peculiar motions (deviations from pure
Hubble flow) of free electrons in ionized regions
\cite{sz80,vishniac87}. There exist other secondary sources of CMB
anisotropies.  However, on angular scales smaller than a few
arc-minutes, the kinematic Sunyaev-Zel'dovich effect is likely to
provide a dominant contribution. For example, it is known that the
temperature anisotropies caused by non-linear growth of density
perturbations, which are often referred to as the Rees-Sciama effect
or integrated Sachs-Wolfe effect, are of order $10^{-7}$ or less
\cite{seljak96}.  These anisotropies depend on the the baryon bulk
physical peculiar velocity, ${\bf v}$, and the number density of free
electrons, $n_{\mathrm e}$.  In our calculations of the anisotropies
we assume that the $\bf v$ is equal to the bulk velocity of the dark
matter and that $n_{\rm e}$ in ionized regions is proportional to the
dark matter density.

The temperature anisotropy $\Theta(\mbox{\boldmath
$\gamma$})=\frac{\Delta T}{T}$ observed in a given line of sight
direction {\boldmath $\gamma$} is (e.g. \pcite{hu99})
\begin{equation}
\Theta(\mbox{\boldmath $\gamma$},\eta_0) =
- \int_{\eta_{\rm rec}}^{\eta_0} \frac{ {\rm d} \eta}{(1+z)} \gamma_i v_{\rm B}^i {\dot \tau},
\label{eq_sol}
\end{equation}
where $\eta \equiv \int (1+z) {\rm d}t$ is conformal time with its
values at recombination and present denoted, respectively, by
$\eta_{\rm rec}$ and $\eta_0$. In eqn.~(\ref{eq_sol}) we have assumed an
optically thin universe. In an optically thick universe these
temperature fluctuations are damped by a factor ${\rm e}^{-\tau}$,
where the optical depth is $\tau=\int {\mathrm d}\eta \sigma_{\mathrm
T} n_{\mathrm e}/(1+z)$, where $\sigma_{\rm T} $ is the cross section
for Thomson scattering.  If the universe became instantaneously fully
ionized after some redshift $z_{\rm i}$, the relation between the
optical depth $\tau(\eta_{\rm i},\eta_0)$ and $z_{\rm i}$ is
approximately obtained as $z_{\rm i} = 100\Omega_0
\left(0.025/\Omega_{\rm b} h\right)^{2/3}
\tau^{1/3}$.  Therefore, if the reionization takes place at $z \ll
100\Omega_0 \left(0.025/\Omega_{\rm b} h\right)^{2/3}$, as is the case
in our reionization model, then the damping factor can be neglected.

The usual procedure to obtain the angular correlation function of
temperature anisotropies in eqn.~(\ref{eq_sol}) is by means of
Limber's equation in Fourier space (see for example
\pcite{peebles80}). However, in this paper, we work in real space
since we have the two point correlation functions of density and
velocity fields directly measured in real space from N-body
simulations.

The temperature angular correlation $C(\theta)$ can be written as
\begin{equation}
C(\theta)  =\sigma^2_{\rm T}
 \int_{{\eta_{\mathrm rec}}}^{\eta_0} {\rm d}\eta
 \int_{{\eta_{\rm rec}}}^{\eta_0} {\rm d}\eta' \gamma_i \gamma_j'
<v^i({\bf x},\eta)  v^j({\bf x}',\eta')
{n_{\rm e}({\bf x},\eta)}{n_{\rm e}({\bf x}',\eta')}> ,
\label{angcorr}
\end{equation}
where $ \gamma_i \gamma'^i=\cos\theta$, and, $\bf x$ and ${\bf x}'$
refer, respectively, to comoving coordinates in the past light
geodesics in the directions {\boldmath $\gamma$} and {\boldmath
$\gamma '$} at $\eta $ and $\eta'$.  We write $n_{\mathrm e}$ terms of
density fluctuations $\delta$ as
\begin{equation}
 n_{\rm e}({\bf x},\eta) = \bar{n}_{\rm e}(\eta)x_{\rm e}
({\bf x},\eta)\left[1+\delta({\bf x},\eta) \right]  ,
\end{equation}
where $\bar{n}_{\rm e}(\eta)$ is the mean total (free and bound)
electron number density at time $\eta$, and $x_{\rm e}({\bf x})$, the
ionization fraction, is unity in ionized regions and zero otherwise.
The correlation lengths of velocity and density fields are small
compared to the Hubble radius so that we can approximate $n_{\mathrm
e} ({\bf x}',\eta') =n_{\mathrm e} ({\bf x}',\eta)$ and similarly for
the $v$, in eqn.~(\ref{angcorr}).

\begin{equation}
\zeta = \left[ {x_{\rm e} \left( 1+\delta\right) \over
\langle x_{\rm e} (1+\delta)\rangle}
 - 1 \right] v_{\rm los},
\end{equation}
where $v_{\rm los}=\gamma_i v^i$ is the velocity component in the
direction $\gamma$. Therefore $C(\theta)$ can be written in terms of
the velocity correlation function $\xi_{vv}(y) \equiv <v_{\rm
los}({\bf x}) v_{\rm los}({\bf x} +{\bf y})> $ and the
density-velocity correlation function $\xi_{\zeta \zeta}(y) \equiv
<\zeta({\bf x}) \zeta({\bf x}+{\bf y})> $, both evaluated for fields
at the same $\eta$.

\begin{equation}
C(\theta) = \sigma_{\rm T}^2
\int_{\eta_{\rm rec}}^{\eta_0} {{\rm d}\eta \over 1+z}
\int_{\eta_{\rm rec}}^{\eta_0} {{\rm d}\eta' \over 1+z'}
\bar{n}_{\rm e}(\eta)\bar{n}_{\rm e}(\eta')
\langle x_{\rm e} (1+\delta)\rangle ^2 \left[\xi_{\zeta\zeta}(|{\bf x}'-{\bf x}|)+
\xi_{vv}(|{\bf x}'-{\bf x}|) + \langle
\zeta({\bf x}) v_{\rm los}({\bf x}')
+\zeta({\bf x}') v_{\rm los}({\bf x})\rangle\right]
.  \label{eq:tri}
\end{equation}
The dominant contribution to $C(\theta)$ is from the term involving
$\xi_{\zeta\zeta}$. The integration over $\xi_{vv}$ yields to phase
cancellation \cite{kaiser84,ostriker86,vishniac87}. The last term in
the integrand also has negligible contribution\footnote{It is
interesting that the contribution from the integral over
$\xi_{\zeta\zeta}$ is still dominant even if we approximate
$\xi_{\zeta\zeta}=\xi_{\delta\delta}\xi_{vv}$, i.e. if we ignore any
correlations between the density and velocity fields}.  We have
checked that the dominant term produces at least an order of magnitude
larger anisotropies than the other terms.

In flat space we use the triangle relation, $|{\bf x}' - {\bf x}|^2 =
x^2 + x'^2 -2 x x'\cos\theta ,$ we first carry out the integration of
eqn.~(\ref{eq:tri}) in terms of $\eta'$ for fixed $\eta$ and
$\theta$. We compute $\xi_{\zeta\zeta}$ at an average redshift $\bar
z$ given by $1/(1+\bar{z}) = \left(1/(1+z_1) + 1/(1+z_2)\right) /2$,
which is an appropriate approximation if the correlation length is
negligible relative to the horizon scale. It is straightforward to
extend the calculation to an open geometry.

From the temperature angular correlation $C(\theta)$, we can obtain
$C_\ell$ as
\begin{equation}
C_\ell =2\pi \int_{-1}^{1}{\rm d}\cos\theta
P_\ell(\cos\theta)C(\theta) ,
\end{equation}
where $P_\ell(\cos\theta)$ is the Legendre polynomial.

\end{document}